## Graphical Table of Contents

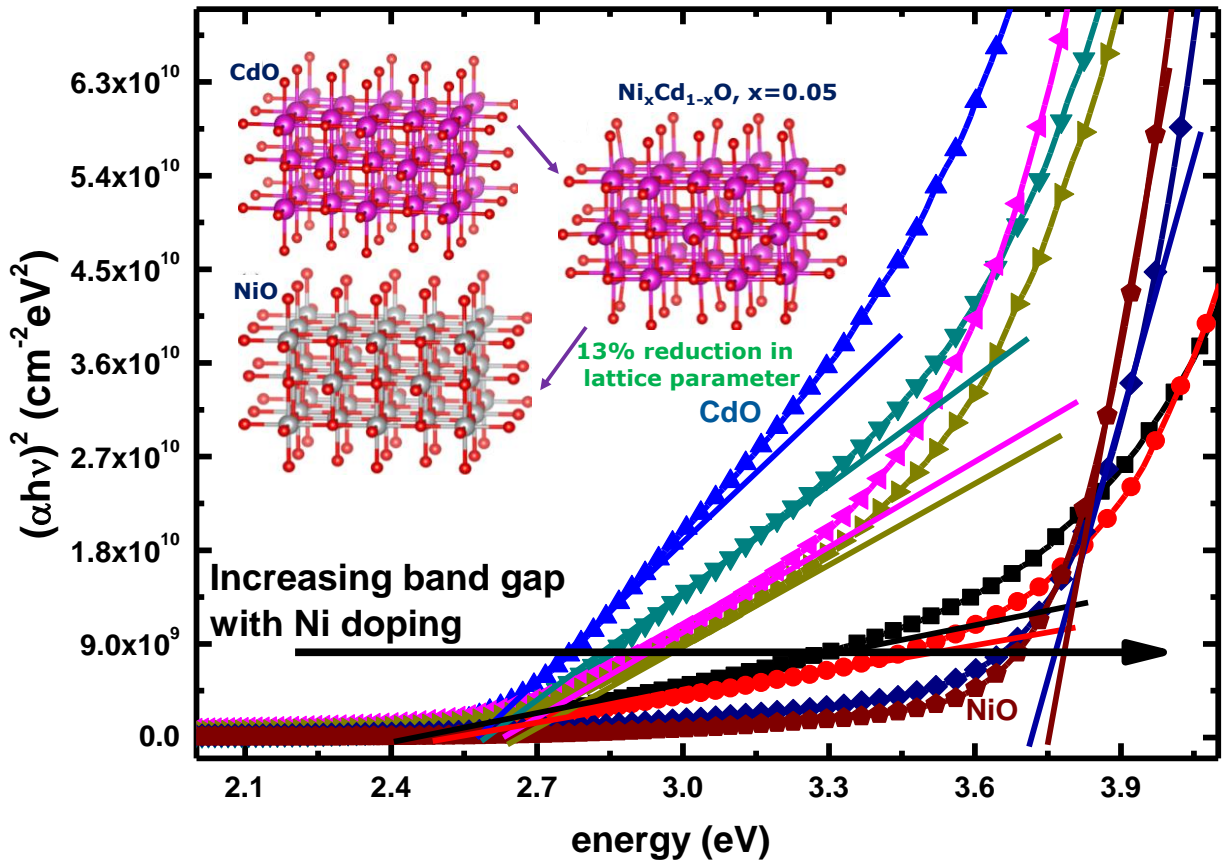

Band gap enhancement and lattice parameter reduction with increasing Ni doping in CdO

# Orbital hybridization induced band offset phenomena in $Ni_xCd_{1-x}O$ thin films

Arkaprava Das[1]*, Deobrat Singh[2], C. P. Saini[1], Rajeev Ahuja[2], Anumeet Kaur[3]

[1]Inter University Accelerator Centre, Aruna Asaf Ali Marg, New Delhi-110067, India
[2]Department of Physics and Astronomy, Condensed Matter Theory Group, Uppsala University, Sweden
[3]Department of Physics, Guru Nanak Dev University, Amritsar, India

**Abstract:**

We present the cationic impurity assisted band offset phenomena in $Ni_xCd_{1-x}O$ (x= 0, 0.02, 0.05, 0.1, 0.2, 0.4, 0.8, 1) thin films and further discussed in the light of orbital hybridization modification. Compositional and structural studies revealed that cationic substitution of $Cd^{2+}$ by $Ni^{2+}$ ions leads to a monotonic shift in (220) diffraction peak, indicating the suppression of lattice distortion while evolution of local strain with increasing Ni concentration mainly associated to the mismatch in electro-negativity of $Cd^{2+}$ and $Ni^{2+}$ ion. In fact, Fermi level pinning towards conduction band minima takes place with increasing Ni concentration at the cost of electronically compensated oxygen vacancies, resulting modification in the distribution of carrier concentration which eventually affects the band edge effective mass of conduction band electrons and further endorses band gap renormalization. Besides that, the appearance of longitudinal optical (LO) mode at 477 $cm^{-1}$ as manifested by Raman spectroscopy also indicate the active involvement of electron-phonon scattering whereas modification in local coordination environment particularly anti-crossing interaction in conjunction with presence of satellite features and shake-up states with Ni doping is confirmed by X-ray absorption near-edge and X-ray photoelectron spectroscopy studies. These results manifest the gradual reduction of orbital hybridization with Ni incorporation, leading to decrement in the band edge effective mass of electron. Finally, molecular dynamics simulation reflects 13% reduction in lattice parameter for NiO thin film as compared to undoped one while projected density of states calculation further supports the experimental observation of reduced orbital hybridization with increasing Ni concentration.

*Author for correspondence:   arkapravadas222@gmail.com (Dr. Arkaprava Das, Research scholar at Inter university accelerator centre)

# 1. Introduction:

The worldwide energy scarcity and environmental issues in the last few decades stimulated the demand of renewable clean energy. Exhaustive efforts have been given for the demand of innovative promising materials which will be having their own intriguing advantages of absorbing the maximum range of the solar spectrum and in turn convert the solar energy into electricity. In recent years, CdO based materials have enticed the scientific community owing to their potential application especially in the fresh unfolding optoelectronic industry [1]. Un-doped CdO is direct band gap semiconductor (~2.2 eV) where the presence of intrinsic anion vacancies generally dictates the *n*-type nature of the materials [1]. In contrast, several theoretical studies revealed that conduction band minimum (CBM) in CdO is located at the centre of the Brillouin Zone ($\Gamma_1$) whereas valance band maximum (VBM) is situated at $L_3$ or $\Sigma_3$ [i.e. away from the zone centre ($\Gamma_{15}$)], indicating the indirect band gap nature of CdO. Therefore, understanding such peculiar band structure and further their role in electrical conduction mechanism is essential in order to design next generation CdO based opto-electronic devices [2]. Moreover, *n*-type conductivity in CdO can be enhanced further by doping of foreign elements (such as Ga, In etc.) which leads to increase the electron concentration beyond $10^{21}$ cm$^{-3}$ [3–5]. Such higher visible range optical transparency along with relatively large conductivity makes it attractive candidate for transparent conductive oxide (TCO) especially for photovoltaic and optoelectronics applications [6]. Furthermore, band gap of CdO can be tuned in desired manner by judicious incorporation of external elements via synthesis of CdO based alloy system though suppression of material conductivity limits the overall device efficiency [7]. For instance, In case of $Cd_xMg_{1-x}O$ alloy, although the band gap increases with Mg concentration, however prominent degradation in electrical transport properties also occur simultaneously [8]. On the other hand, Cd rich $Cd_xZn_{1-x}O$ alloys with rocksalt structure shows an increasing band gap with increasing Cd percentage without significant degradation in transport properties [9]. In fact, doping of Mg, Zn in CdO definitely results into a widening of band gap and explained in the light of critical light absorption model which is accompanied by both direct and indirect transition. For optoelectronic applications, it is highly preferable to develop an alloy which not only exhibit both *p*-type and/or *n*-type electrical behaviour but also will play a significant role as band gap enhancing agent. In this context, Nickel Oxide (NiO) is a wide band gap semiconductor (~3.7 eV) [10] and also a prominent antiferromagnetic insulator with a few paramount applications such as in rechargeable batteries, electrochromic films, catalysis, giant magnetoresistive (GMR) spin valve structure, fuel cells and gas sensors etc. [11–14]. The VBM, in NiO, is likely to be located near the Fermi stabilization energy level ($E_{FS}$) [15], resulting the *p*-type conductivity and therefore makes it distinguishable among the other metal oxides [16–18]. Such *p*-type nature in NiO is mainly attributed to the presence of Nickel vacancies, which evolved inherently [19]. Moreover, the electronic transition in NiO which causes the band gap, takes place between the localized *d* states rather than between band like states like usual semiconductors [20]. So the term "optical gap" instead of "band gap" will be better to be casted off.

In the present report, we have synthesized the $Ni_xCd_{1-x}O$ (NDO) thin films for the whole composition range and a detailed study has been carried out on their structural, optical and chemical properties. In particular, we demonstrate the doping induced local strain, electron-electron, electron-ionized impurity interaction and strong electron phonon coupling leads to renormalization of band gap in NDO thin films. The dramatic dependence of the optical gap, moreover, with composition also has a strong correlation with phonon scattering such as optical phonon deformation potential, charged impurity scattering and also phonon stiffening as evidenced by Raman spectroscopy. Moreover, we consider the *p-d* hybridization between O *p* and Ni/Cd *d* orbitals in order to determine the CBM and VBM energy levels, suggesting that *p-d* hybridization becomes less effective with increasing Ni doping and subsequently affects band gap renormalization by reducing the effective mass of electron at conduction band edge. Complementary X-ray absorption near-edge spectroscopy (XANES) has been investigated, showing clear change in electronic spectra and local coordination environment in O *K*-edge with increasing Ni doping. Furthermore, increasing peak area and intensity of satellite features with increasing Ni doping has been demonstrated by X-ray photoelectron spectroscopy (XPS), suggesting the evolution of new phases with doping. Subsequently, molecular dynamics simulation along with projected density of states calculation reflects reduction in lattice parameter and orbital splitting with increasing Ni doping respectively. From PDOS calculation, this has been observed that allowed band of Fermi level pinning with respect to the vacuum level is observed to be 5.5 eV from CdO to NiO composition. This wide change of Fermi level has been explained with the help of Amphoteric Defect Model (ADM) [21] using the concept of localized *d* shell defects.

## 2.1 Experimental:

Undoped and doped CdO films were deposited on the corning glass and silicon substrate using sol-gel spin coating technique. Details of the sample preparation, structural, optical, microscopic, Raman spectroscopic measurements are reported elsewhere [22].

Different instrumental facilities are utilized for structural, optical, spectroscopic and electronic structure (oxygen *K*-edge) characterization. The details of the concerned instruments are described elsewhere [22]. Surface chemical analysis has been carried out using an Al-$K_\alpha$ source (incident energy = 1486.7 eV) in an Omricon nanotechnology XPS system (Oxford instruments, ESCA+ model) integrated with 124-mm hemispherical electron analyser. Calibration of binding energy scale was made using carbon 1s edge.

## 2.2 Computational Method

In the present work, the first-principles calculations have been performed within the framework of spin-polarized density functional theory (DFT) [23,24] as employed in Vienna Ab initio Simulation Package (VASP) [25,26]. The meta-generalized gradient approximation (MGGA) [27] with Perdew, Burke, and Ernzerhof (PBE) parameterization is applied for electronic exchange and correlation functional [28]. A 15×15×15 Γ-centered k-mesh was sampled to optimized the unit cell of rock-salt of CdO and NiO and the kinetic energy cut-off of the plane wave bases is set at 1200 eV (CdO) and 500 (NiO) for unit cell of both structures. While in case of $Ni_xCd_{1-x}O$ (x≈0.05, 0.10, 0.2, 0.4, 0.8), the kinetic energy cut-off of the plane wave bases is set at 800 eV during the calculations. For the doping of Ni atoms in rock-salt CdO structure, we have taken (2×2×1) supercell. The atomic coordinates are fully relaxed until the atomic force reached upto $10^{-3}$ eV/ Å and the total energy convergence criterion between consecutive self-consistence field (scf) cycles is set as $10^{-6}$ eV.

The valence-electron configurations for the elements of rock-salt $Ni_xCd_{1-x}O$ are Ni $3d^8 4s^2$, Cd $4d^{10}5s^2$ and O $2s^2 2p^4$. Due to the presence of d-electrons in Ni and Cd atoms, we have employed on-site Coulomb correction within GGA+U [29], to predict realistic properties for these materials. In this study, the electronic interactions are described within the GGA+U formalism for $Ni_xCd_{1-x}O$, where on-site Coulomb corrections are applied and GGA+U approach yields an acceptable electronic band structure, for rock-salt $Ni_xCd_{1-x}O$ (x≈0.0, 0.05, 0.10, 0.2, 0.4, 0.8, 1.0). The on-site Coulomb corrections are applied on the $3d$ and $4d$ orbital electrons of Ni and Cd ions (U-d) for $Ni_xCd_{1-x}O$, respectively. In the present work, we have used the GGA+U methods for on-site Coulomb corrections $U_{eff} = U−J$, where U is the Hubbard parameter and J is the exchange parameter. During the whole calculations, we utilized U = 12.0 eV (Cd), 8.0 eV (Ni), and J = 0.0 eV (Cd), 0.90 eV (Ni) for correlation effect of localized $d$-states for Cd and Ni atoms [30–32].

## 3. Result and discussion:
### 3.1 Compositional and structural studies:

Compositional analysis has been done using Rutherford backscattering (RBS) technique while the simulation is performed using Rump software. The samples doped with 0%, 3%, 5%, 10%, 20%, 40%, 80%, 100% of Ni are abbreviated by 4Cd, 3% Ni, 5% Ni, 10% Ni, 20% Ni, 40% Ni, 80% Ni, 100% Ni, respectively. Figure 1 (a), (b), (c) and (d) show the recorded RBS spectra with simulation for undoped 40%, 80%, 100% Ni doped samples whereas film thickness, extracted by fitting of spectra is summarized in Table 1, respectively. Moreover, compositional percentages of constituent elements are also shown in Table (S1) (Electronic Supplementary Material [ESM]), clearly indicating the gradual reduction in the Cd concentration with increasing Ni percentage. On the other hand, increase in Ni doping also leads to decrease in O concentration and therefore film become less O deficient in nature. It is also noteworthy to say that with enhancing Ni dopant, the amount of linear diffusion inside the Si substrate provided for simulation diminishes monotonously. This gives a direct signature that diffusivity of Cd is more as compared to Ni. In case of 100% Ni sample, no linear diffusion has been provided inside the Si substrate for the fitting of RBS spectra. Unlike 100% Ni one, we observe the high backscattering yield in the 4Cd RBS spectrum owing to relatively higher atomic mass and backscattering cross-section of Cd than Ni particularly for $165^0$ backscattering geometry [33].

Figure 2 shows the glancing angle X-ray diffraction (GAXRD) pattern for 4Cd, 3% Ni, 5% Ni, 10% Ni, 20% Ni, 40% Ni, 80% Ni, 100% Ni thin films, showing the cubic structure (JCPDS: 78-0653) of CdO films though polycrystalline nature was evident up to 40% Ni doping. Close inspection further revealed that the peak intensity corresponding to (111) diffraction plane decreases with FWHM broadening monotonously with increasing Ni percentage, suggesting that the reduction in average particle diameter with Ni concentration. It is well known fact that ionic radii of $Ni^{2+}$ ion (~0.63 Å) is significantly smaller than that of the $Cd^{2+}$ ion (~0.97 Å) [34] and therefore, the generation of lattice distortion due to mismatch in ionic radii can be quantified by Goldschimdt's Tolerance Factor (GTF) which is given by the following relation [35].

$$t = \frac{r_A + r_O}{\sqrt{2}(r_B + r_O)} \tag{1}$$

Where $r_A$, $r_O$, $r_B$ are the ionic radii of $Cd^{2+}$, $O^{2-}$, $Cd^{2+}/Ni^{2+}$ and $r_B = (1-x)r_{Cd} + xr_{Ni}$, respectively. The variation in GTF factor with Ni doping is depicted in Figure 3 whereas the calculated values are summarized in Table 1, respectively. In fact, the gradual increments in GTF factor with increase in Ni concentration reflect the instability suppression via generation of stress, consistent with our GAXRD results.

Moreover, the decrease in average crystallite size with increasing Ni concentration gives a direct evidence of substitutional replacement of $Cd^{2+}$ ion by $Ni^{2+}$ ion which in turn leads to alter the stoichiometry of CdO film. By using the (111) peak, the average crystalline size of doped and un-doped CdO was calculated by Scherrer's equation [36] and outcome was summarized in Table 1.

$$D = \frac{0.9 \times \lambda}{\beta \cos\theta} \tag{2}$$

Where D is crystallite size, $\lambda$ is the x-ray wavelength, $\theta$ is the Bragg's diffraction angle and $\beta$ is full width at half maximum.

In addition, the average crystalline size for (220) diffraction peak is also calculated and incorporated in Table (S1) (ESM). Close inspection further revealed that both (111) and (220) diffraction peaks moves towards the higher 2θ value with increase in Ni concentration though the movement in former one is more pronounced than later one. Such monotonic shift in 2θ value with increasing Ni concentration is attributed to the reduction of lattice parameter and evolution of compressional stress inside the lattice [37]. In fact, this local stress is mainly originated due to presence of defects like Ni interstitials and/or Ni clusters and so getting more prominent with increasing Ni dominance. Beside this, the substitution of $Cd^{2+}$ by $Ni^{2+}$ ion also leads to modify the short range order parameter owing to prominent difference in the electronegativity between Cd (~1.69) and Ni (~1.91) which likely to be another possible reason for producing such local micro structural strain. Since no significant difference in thermal expansion coefficient of CdO (14e-6 $K^{-1}$) and NiO (1.39e-6 $K^{-1}$) in wide temperature range (~300 to ~800 K) is observed previously, we believe electronegativity mismatch could be the main possible reason behind this strain which in turn results the band gap widening (will be discussed in the following). Moreover, no XRD peaks for 80% Ni and 100% Ni samples were detected, indicating that poly-crystalline films got amorphized at that higher level of Ni doping and thus limit to calculate the average particle diameter of them. Unlike L. Gao *et al*. works where 3% and 5% Ni doped CdO ceramic exhibit a secondary NiO phase, [34] here, no signature of secondary NiO phase was observed in GAXRD pattern even after 40% Ni doping, suggesting that doping in ceramic bulk samples using conventional solid state reaction method is completely different approach then thin films prepared by solgel technique.

### 3.2 Topographical study:

In order to explore in change in surface morphology and roughness with increasing Ni doping, detailed AFM study has been carried out in tapping mode. Here, Figure 4(a), 4(b), 4(c), 4(d) shows the AFM images for 4Cd, 5% Ni, 20% Ni, 80% Ni samples, while corresponding root mean square (*rms*) roughnesses are summarized in Table 1, respectively. From the images, it is clear that grown films are consistent and dense in nature without any huge crack where interconnected grains are homogeneously distributed all over the surface without any void formation. From the Table 1, it is quite clear that roughness decreases and grains are becoming finer with increasing Ni doping up to 20% which is a direct evidence of reducing crystallinity with increase in Ni concentration, matches well with our XRD results. Such nano-grain refinement effect may be attributed to the immobilizing effect of Ni additive which restrain the grain boundary movement and thus the grain growth [38]. Synonymous grain refinement phenomenon has also been reported for Ni doped ZnO ceramics [39]. This nano-grain refinement effect increases the overall grain boundary area which in turn leads to increase grain boundary scattering and subsequently will increase electrical resistivity of the film [40]. The smaller grain size causes a lower surface roughness and consequently results in reduction of the propagation loss for surface acoustic wave (SAW) which might lead to an enhanced efficiency in the photovoltaic devices [41] though the impact on the propagation of SAW is modest as the variation in *rms* roughness is in nano-scale level. Here, the chemical composition driven micro structural modification can significantly influence the band gap widening phenomena [42].

### 3.3 Study of optical gap modifications:

Figure 5(a) shows the optical transmittance spectra for NDO thin films with different Ni concentration, showing the more than 60% transparency for all samples particularly in the wavelength range of ~600 to ~800 nm. In fact, recorded spectra of the films grown on a smooth substrate surface don't consist of interference fringes and further investigated by Pankove analysis [43]. Close inspection further revealed that visible range transmittance increases from ~55% to ~85% whereas the fundamental absorption edge is found to get blue shifted and become sharper with increasing Ni concentration. Such blue shift of the cut off wavelength gives an evidence of band gap widening with increasing Ni doping. The calculated values of the refractive index for all the samples are summarized in Table 1. The corresponding absorption spectra before and after Ni doping are also shown in Figure 5(b). The absorption coefficient α is related with the photon energy (hν) by the following relation [44].

$$\alpha h\nu = A(h\nu - E_g)^n \tag{3}$$

Where, A is a constant, hν is the incident photon energy, $E_g$ is the optical band gap. Since the CdO is direct band gap material [45], n is consider to be equal to 2 in the equation for calculating the band gap of the films. Therefore, the extrapolation of the linear plot provides the value of band gap for direct allowed transitions. As can be seen from Figure 5(b), Tauc spectrum for Pure CdO thin film encompasses a broad shoulder at the lower energy side and thus the signature of Cd rich phase. On the other hand, the broad shoulder starts to get vanish with increasing Ni density and absorption edge starts to shift towards higher photon energy. In particular, the spectra for higher Ni concentration such as 80% Ni and 100% Ni thin films exhibits a sharp rising steep which signifies Ni dominance [46]. In Tauc relation, moreover, it is assumed that conduction band and valence band are parabolic though it's different in actual scenario for CdO. Therefore, owing to such highly non-parabolic nature of conduction band in CdO, it is difficult to follow the Tauc relation in order to determine optical bang gap correctly [47]. The high non-parabolicity of the conduction band, in fact, is the consequence of additional donor states due to higher carrier concentration in CdO. Generally the semiconductor having large electron concentration, the absorption edge is displaced by Burstein-Moss Shift (BMS) [32, 33]. It is clearly reported by Francis *et al.* [37] that with increasing Ni doping electron concentration decrease significantly though the band gap values increase gradually. So filling of lower sub-states of the conduction band by the excess electrons (BMS effect) can't explain this band gap widening phenomenon as electron concentration is decreasing with increasing optical gap here. Therefore, the observed band gap widening phenomena can be explained in the light of band gap renormalization effect including electron-electron and electron–ion interactions. Generally the renormalization effect is prominent for material having large electron concentration. However, with increasing Ni percentage conduction band edge (CBE) moves towards the $E_{FS}$ and as a consequence the formation energy for acceptor and donor like native defects become equal and in turn results a remarkable reduction of the mobility and electron concentration [50]. Indirectly one can enunciate that reduced concentration of electrically active donor impurities might be corroborated to the formation of compensating acceptor type native defects [15]. Therefore, with increasing Ni dopant the effect of electron-electron and electron-ion interaction will become more prominent in band gap widening phenomena. The shift of unperturbed intrinsic band gap can be subdivided in two parts [51].

$$\Delta E_g = E_g - E_g^0 = (\Delta E_g)_1 + (\Delta E_g)_2 \tag{4}$$

$\Delta E_g$ is a positive quantity and $E_g^0$ is the unperturbed intrinsic optical gap. $(\Delta E_g)_1$ is the shift in the optical gap due to local strain in the presence of impurity centres. Here $(\Delta E_g)_1$ can be associated to local strain, and already been discussed in GAXRD results. With increasing Ni percentage the strain as well as the band gap is enhancing which gives direct evidence that pressure coefficient of the band gap is having a positive value. Thus, optical band gap enhancement phenomena will try to fetch the alloy system to a stressed state.

$(\Delta E_g)_2$ is the shift considering Coulomb interaction among the electrons plus their interaction with the ionized impurity. It also includes the BMS effect. Therefore

$$(\Delta E_g)_2 = E_{BM}(n) - E_{el-el}(n) - E_{el-i}(n) \quad \text{[ref. 35]} \tag{5}$$

The shift in conduction-band edge due to the electron-electron interaction is given by the following equation[51]

$$E_{el-el}(n) = -\frac{2e^2 k_F}{\pi \epsilon_0} - \frac{e^2 \lambda}{2\epsilon_0}\left[1 - \frac{4}{\pi} tan^{-1}\left(\frac{k_F}{\lambda}\right)\right] \tag{6}$$

Where $k_F$ and $\lambda$ are the Fermi wave vector and Thomas Fermi screening parameter given by the following relations

$$k_F^3 = 3\pi^2 n \tag{7}$$

Where n is the concentration of free charge carrier

$$\lambda = \frac{2e}{\hbar \left(\frac{3n}{\pi}\right)^{\frac{1}{6}} \left(\frac{m^*}{\epsilon_0}\right)^{\frac{1}{2}}} \tag{8}$$

Where $m^*$ is the effective mass of electron at the Fermi energy, and $\epsilon_0$ is the static dielectric constant.
The other contribution for the conduction band shift is due to electron-ionized impurity scattering, given by following relation.

$$E_{el-i}(n) = -\frac{4\pi n e^2}{\epsilon_0 a_0 \lambda^3} = -\frac{e\hbar}{2}\left[\frac{\pi^3 n}{\epsilon_0 m^*}\right]^{\frac{1}{2}} \tag{9}$$

Where $a_0 = \frac{\epsilon_0 \hbar^2}{m^* e^2}$ is effective Bohr radius.

The value of the effective mass in equation (7) and (8) depends upon carrier density, the position of Fermi level and also on the carrier concentration. Usually, Fermi level locates above the conduction band minima for n type degenerate semiconductor like CdO. The value of this energy dependent effective mass has been calculated using two-band k.p method and incorporating the effect of band gap renormalization reported by Berggren and Sernelius [51].

For a given Fermi level ($E_F$) carrier density, band dispersion and density of states integrated effective mass were determined by using the following two equations.

$$n(E_F) = \int g(E)f(E,E_F)dE \qquad (10)$$

And

$$m_{av}^*(E_F) = \frac{1}{n(E_F)} \int m^*(E)g(E)f(E,E_F)dE \qquad (11)$$

Where $f(E,E_F)$ is the Fermi-Dirac distribution, $g(E)$ is the density of states and $m^*(E)$ is the energy dependent effective mass of electron.

In previous study, Jefferson *et al.* [6] reported that decrease in electron concentration leads to suppression in $m_{av}^*(E_F)$ monotonously. Since the formation of oxygen vacancies enhance the electron concentration inside the oxide system, reducing electron concentration, here, can be explained by decreasing oxygen vacancy population with increasing substitutional impurity concentration. For n-type material $E_F > E_g/2$ the formation energy for $Ni_{Cd}$ ($Ni^{2+}$ substituted $Cd^{2+}$) is lesser than formation energy for oxygen vacancy. In order to maintain overall charge neutrality of the system, increasing Ni concentration drives the Fermi level towards the CBM at the expenses of electronically compensated oxygen vacancies, [52,53] resulting modification in the distribution of carrier concentration which eventually affects the band edge effective mass of conduction band electrons. So with increasing Ni dopant, the effective mass will decrease as carrier concentration goes to a lower value. The proportional relation of average effective mass of electron with carrier concentration has also been endorsed by Dou *et al.* [54] via the following relation.

$$m^* = m_0^* + cn_0 \qquad (12)$$

Where $c \sim 10^{-20} m_0$ cm$^{-3}$, $m_0^* = 0.121 m_0$, $m_0$ is the effective mass of an electron. However, in maximum cases we found a larger value of the band edge effective mass of electron. It is redundant to say that it is the consequence of the assumption of parabolic model, implicit in the calculations [55].

Due to this decrement in effective mass the overall magnitude $E_{el-el}(n)$ and $E_{el-i}(n)$ increases as effective mass resides in the denominator in the equations (5) and (8) respectively. The increment in the magnitude will result an increment in the overall electron-electron, electron-ion interaction and subsequently there is an upward shift of the conduction band minimum (CBM). Both the equation (5) and (8) contains the dielectric constant of the host matrix. The static dielectric constants for CdO and NiO are 18 [56] and 12 [13] respectively. So with increasing Ni composition in NDO alloy static dielectric constant will also get change and subsequently it will also change the impact of electron-electron and electron-ion interaction.

The unusual composite dependence NDO alloy can be fitted via second order non-linear bowing equation[57,58].

$$E_g^{NiCdO}(x) = E_g^{NiO}(x) + E_g^{NiO}(1-x) - b_{opt}x(1-x) \qquad (13)$$

Where, $E_g^{NiCdO}(x)$ is optical gap of the composite alloy and $b_{opt}$ is the optical bowing parameter. Bernard and Zunger [57,58] have mentioned that $b_{opt}$ consists of three major contribution. Firstly, the volume deformation (VD) due to change in the lattice dimension (XDR pattern gives us the signature of lattice deformation), secondly charge exchange (CE) in the alloy relative due to difference in the electronegativity of Cd and Ni atom and thirdly structural contribution (S) caused by cation-anion bond length relaxation in the alloy [59]. Therefore,

$$b_{opt} = b_{VD} + b_{CE} + b_S \qquad (14)$$

For NDO alloy each of these three contributions in $b_{opt}$ would be prominent. The second order fitting yield $b_{opt} = 0.19$ eV. However, from Figure 6 we observe the fitting is rather poor and so any further comment can't be made regarding this fitting.

Mainly the theoretical model in n type material where electron gas occupy the conduction band sub-states is (BMS effect) considerable for the electron concentration higher than Mott critical density, (electron concentration)$^{1/3}$*($a_H$) ≈ 0.25 ( $a_H$ is the effective Bohr radius) [51].

Generally for CdO the main reason for emanating donor states are the Cd interstitial or Oxygen vacancies [47]. Surface electron accumulation is a well known phenomena for CdO thin films [60]. So here position of the surface Fermi level with respect to conduction band minimum (CBM) can be described in terms of Burstein-Moss shift [32, 33].

$$\Delta E_{BM} = E_f - E_C = \frac{\hbar^2}{2m^*}(3\pi^2 n)^{\frac{2}{3}} \qquad (15)$$

Where $m^*$ is electron effective mass and $n$ is the electron concentration. With increasing Ni doping electron concentration reduces for NDO alloy and therefore, both Fermi level will move downwards while CBM will shift upward. Due to overall alignment between surface and bulk region, Fermi level will go downwards i.e. towards conduction band minimum. So from this discussion we can conclude that shifting of conduction band will influence the optical gap modification more rather that valance band shifting. This shift will influence $n(E_F)$ in equation (10) and subsequently also the band edge effective mass of the electron.

In a nutshell, we can say that the reduction of free carriers with increasing Ni doping results a weakening of mutual exchange and coulomb interaction. Subsequently upward shift of CBM have been observed which is accentuated by repulsive interaction between extended, unoccupied conduction band and occupied donor states (such as $d$ levels) of Ni [61]. Beside this, such reduced electron concentration also influences the valance band states as well. In fact, increasing Ni concentration in CdO reduces the propensity of the material of being n type by reduction of free electron concentration though effective hole concentration also reduces simultaneously in order to maintain overall charge neutrality. Consequently, mutual exchange is replaced by a less dynamically screened interaction and this could be the possible reason for upward shifting of the valence band maxima (VBM). Francis *et al*. [37] have also been discussed such upward shifting by using an energy band diagram however, first principle studies, on the other hand, suggest that optical gap modification mainly can only occurs due to the shifting of CBM than VBM [62].

In this scenario Bohr excitonic radius (BER) of CdO is not reported earlier[63] though clearly mentioned by Lopez-Ponceet *et al*. [64] that BER increases for Cd doped ZnO alloy with increasing Cd concentration, indicating the relatively larger BER in CdO than ZnO [65]. Similarly, in case of NiO, however, quantitative information available regarding BER is scarce but quantitatively it is reported that BER of NiO is ~0.13 nm which is even further lower than ZnO [66,67]. Therefore, we can anticipate that increasing Ni doping in CdO will decrease the BER of NDO alloy and particle diameters for all the samples are below ~14 nm, suggesting that quantum confinement effect may be another factor for optical gap enhancement.

### 3.4 Micro-Raman studies

CdO rocksalt structure exists in $Fm\bar{3}m$ space group symmetry. It is well established that A1, E1 both are Raman and IR active branches, there symmetries are polar, doubly degenerated and split into TO (Transverse Optical) and LO (Longitudinal Optical) components with different frequencies. $E_2$ (High) and $E_2$ (Low) branches are non-polar in nature and so both Raman as well as IR inactive. Figure 7 shows the main Raman spectral features of pure and NDO thin films. For CdO molecules only second order Raman scattering is allowed and the excitation wavelength used in the measurement was 514.5 nm which is near to the optical gap value for pure CdO. Therefore the probability of first order Raman spectra can't be ruled out completely. For 4Cd thin film a peak at ~477 $cm^{-1}$ is detected in the spectra. It has been reported theoretically that CdO molecules have LO active phonon modes at ~478 (25) $cm^{-1}$, ~952 $cm^{-1}$ and TO active phonon mode at 262 (3) $cm^{-1}$ [55, 56]. In our present spectra, we can only clearly observe the presence of ~477 $cm^{-1}$ LO phonon mode. From the perspective of Raman selection rule LO and TO phonon modes are dipole forbidden so all the features for 4Cd thin film can be attributed to the second order Raman scattering process. For all the thin films prominent presence of LO mode is quite obvious as incident excitation laser light was perpendicular to the sample surface. From the retrospective discussion, we know that in NDO thin films $Cd^{2+}$ ion is substituted by $Ni^{2+}$ ion which breaks the translational symmetry in CdO lattice. Therefore, the phonons having wave vector away from the Brillouin zone centre start to contribute in Raman scattering phenomena inside the host matrix and $q = 0$ ($q$ is wave vector) selection rule relaxes [70]. As a consequence due to Ni doping asymmetric broadening and blue shift of the 477 cm$^{-1}$ LO phonon mode has been observed [71]. We observe 19 cm$^{-1}$ blue shift up to 80% Ni doped thin film with respect to 4Cd thin film. LO phonon modes in CdO molecule exhibit an atomic displacement along c axis. The frequency shift is proportionally related with the short range order parameter along c axis which is nothing but the distance between Cd and O atom. After Ni doping certainly the short range order parameter as well as the bond length will get change caused by the prominent mismatch in electronegativity between $Cd^{2+}$ and $Ni^{2+}$ ion. As a result lattice distortion and modulation of the wave function will take place via deformation potential scattering process [72]. The inter-atomic distance $d$ between the Cd atoms in CdO (3.32 Å) [2] is much larger than the distance between Ni atoms in NiO (2.97 Å) [73]. So it is expected that with increasing Ni doping the inter-atomic distance in the second coordination will keep on decreasing. In the tight-binding band structure calculations of Harrisons scheme [74], the matrix element of the first neighbouring interaction is scaled to $d^{-2}$. Therefore, with increasing Ni doping this interaction will get more prominent as the substitution of $Ni^{2+}$ at the $Cd^{2+}$ site produce structural disorder and results phonon stiffening. So due to this complex interaction mechanism we can speculate the 19 cm$^{-1}$ phonon stiffening (blue shift) in the 477 cm$^{-1}$ LO mode with increasing Ni doping. This phonon stiffening also reconfirms the retrospective discussion of the XRD pattern about compressive strain with ionic substitution. However, for the 100% Ni thin film spectral features changes drastically which gives direct evidence that the presence of Cd in NDO alloy is very much robust as even up to 80% Ni thin film spectral features are quite synonymous. For 100% Ni thin film all optical phonon modes between 400 cm$^{-1}$ to 600 cm$^{-1}$ arise from first order Raman scattering. LO, 2TO and 2LO modes are evolved at 570 cm$^{-1}$, 800 cm$^{-1}$ and at 1100 cm$^{-1}$ respectively. A weaker sub peak is also evolved at 963 cm$^{-1}$ (LO + TO) [75]. The 2TO phonon mode is found to evolve after Ni doping and gets strongest for 100% Ni thin film. However after 5% Ni doping this 2TO peak shows small amount of phonon stiffening with further Ni doping. Theoretically reported force constant value of pure CdO is 101.57 N-m$^{-1}$ [76] but due to lack of experimental results we can't cross verify this value with existing experimental literature. However, it is reported for Ni doped ZnO that with increasing Ni doping particle size reduces, surface phonon modes shows a phonon stiffening and increment in force constant [77]. In present case also we observe stiffening of LO phonon mode. The calculated reduced masses of NDO alloy are cited in Table (S1) which is also observed to be decreasing like Ni doped ZnO system [77] with increasing Ni percentage. Phonon scattering is also a grain boundary area dependent phenomenon. From AFM images we observe that fractional grain boundary area (dislocation) is increasing with increasing Ni doping [78]. Therefore phonon scattering from grain boundary will increase which will decrease the mobility as reported by Francis *et al.* [37].

All Raman spectra have been recorded for the same exposure time of the exciting electromagnetic radiation. The drastic intensity enhancement of 477 cm$^{-1}$ LO mode with Ni doping is deeply correlated with the aforementioned ionic substitution. Eventually this intensity enhancement is the result of strong electron phonon coupling [35]. Therefore, with decreasing particle size and electron concentration it is essential to involve the electron-phonon interaction in explaining the band gap renormalization phenomena. For lower carrier density (lower than Mott Critical density) band gap widening is expressed considering the following effects, [79]

$$\Delta E_g = E_{ex}^0 + E_c + E_{e-ph} \qquad (16)$$

Where, $E_{ex}^0$, $E_c$, $E_{e-ph}$ are the exchange, correlation and electron-phonon interaction self energy. The degree of disorder with increasing Ni percentage enhances the interaction between charges transferred between $Cd^{2+}$ and $Ni^{2+}$ and lattice distortion. This results an enhancement in electron-phonon coupling which might lead to the intensity enhancement of 477 cm$^{-1}$ LO phonon mode. However, the electron-LO phonon coupling also strongly depends on the free carrier concentration. The aforementioned reduced oxygen vacancy concentration on the surface of nanoparticles can also change both the short range and long range Coulomb forces which might be another possible reason for 19 cm$^{-1}$ phonon stiffening in LO phonon mode. Generally, less number of high frequency phonons are deeply influenced by point defect scattering. At higher doping level the overall mobility of a system is expressed by the following equation, [80]

$$\mu^{-1} = \mu_{cc}^{-1} + \mu_{ph}^{-1} \qquad (17)$$

Where $\mu_{ph}$, $\mu_{cc}$ are the room temperature phonon mobility and the mobility of electrons scattered by charged centres. Therefore at higher doping level overall mobility is mainly influenced by phonons. When the Ni concentration is high enough it will scatter the room temperature thermally activated phonons via optical phonon deformation potential and charged impurity scattering [53]. So ionic substitution with Ni doping will aggravate the electron-LO phonon coupling via optical phonon deformation potential which results an intensity enhancement of 477 cm$^{-1}$ LO phonon mode and also plays a significant role in renormalization of band gap [81]. For lower carrier concentration (electrons in the conduction band and holes in the valance band) with a lesser dynamically screening, the free carriers renormalizes the band gap via gradually escalating electron-LO phonon coupling [82].

**3.5 Soft X-ray absorption spectroscopy:**

Oxygen $k$ edge XANES for 4Cd, 5% Ni, 10% Ni, 80% Ni, 100% Ni thin films (Figure 8) are recorded in TEY (Total Electron Yield) mode with normalized μ(E) versus photon energy (eV). As discerned for spectra, peaks labelled as features A', B' are clearly observed in pure CdO thin film in figure. These prime features are generated due to dipole allowed transitions between O 1$s$ core state to unoccupied *p* core states above Fermi level [83]. If bonding between $O^{2-}$ and $Cd^{2+}$ would have been ionic then *p* core states would be occupied and aforementioned dipole allowed transitions would become forbidden. Therefore, the bonding is not purely ionic rather of mixed character [84]. With increasing Ni doping not only the nature of A', B' features change but also the signal generated due to interference effect of multiple scattering signals beyond 550 eV photon energy gets modified, respectively. From the X-ray results, we have concluded that replacement of $Cd^{2+}$ with $Ni^{2+}$ ion takes place which certainly going to modify the bond length due to significant difference in electronegativity and ionic radii and also the crystal field which might be the possible reason in modification of multiple scattering signals. It is well reported phenomena that a strong orbital hybridization i.e. *p-d* hybridization always takes place in rocksalt IIB-VIA compounds which becomes the origin of anomalous valence band structure [85]. As CdO falls in that domain, therefore strong *p-d* hybridization is quite expected for undoped films. However, it is also well reported by Demchemko *et al.* [84] that angular momentum projected local density of states provides distinct indication conduction band minimum substates are constructed of Cd 5*s* states hybridized with O 2*p* states. In contrast, NiO valence band shows oxygen 2*p* character and conduction band predominantly shows Ni 3*d* character [86] which is in good agreement with our performed density of states calculations. Therefore, with increasing Ni doping certainly there will be a significant change in orbital hybridization, localized and extended sub states of in CdO host matrix. From the XANES spectra, one can observe clearly that with increasing Ni doping the peak B' gets flattened when the concentration reaches up to 80%, evolution of A, B, C, D peaks takes place which are completely different from undoped and lower doped ones. As can be shown by spectra for 80% Ni and 100% Ni samples, the A, B, C and D peaks are getting more and more prominent with increasing Ni content where spectra is mainly attributed to the transition to unoccupied *p* character which is mixed in conduction band. The peaks designated by the B (543 eV) and C (547.3 eV) as shown by an arrow are attributed the Ni $e_g$ sub-band generated due to the orbital hybridization between Ni 3*d*-O 2*p* orbitals [87]. In fact these bumps are related to Ni 4*p* bands whereas bumps indicted by D (553.4 eV) are associated with Ni 4*s* band [87]. Like CdO, NiO also exists in rocksalt structure and Ni ions are coordinated to perfectly ligand octahedral and in turn results the splitting of Ni 3d orbitals into $t_{2g}$ and $e_g$ levels [88]. The bump designated by A (538.7 eV) corresponds to $t_{2g}$ level. When Ni is doped in CdO host matrix, the unusual composition dependence of band gap can be attributed to the partially filled *d* shell of Ni on the electronic band structure [7]. When a TM atom is incorporated inside a semiconductor, it can either act as a donor by leaving one *d* electron or can act as an acceptor by acquiring one electron form the valence band of the host matrix. The noteworthy features of such kind of *d* donor and *d* acceptor sates are independent of the host semiconductor matrix as their charge transition energy does not vary at absolute scale with respect to the vacuum level [61]. As a consequence of this, highly localized states inside or at the neighbourhood of conduction band are introduced in CdO host matrix. Therefore, more will be the doping, more would be the density of these localized states and subsequently more will be interaction of those localized states with extended states. $Ni_xCd_{1-x}O$ system will act as a highly mismatched alloy where anti-crossing interaction (AC) between highly localized and extended states will dictate the electronic band structure i.e. nothing but the interaction between Ni *d* levels and extended conduction band states of CdO [89]. The strength such of kind of AC is dependent upon energy difference between conduction band edge and localized states [90]. Usually due to this anti-crossing interaction, conduction band splits into two sub bands. The deconvolution of A' and B' features are quite clearly observed with increasing Ni doping percentage. The aforementioned anti-cross interaction might be the possible reason behind such kind of deconvolution though distortion due to crystal field is also another significant factor for such kind of $t_{2g}$ and $e_g$ sub-bands splitting [91]. The cubic field splitting results in splitting of five Ni 3*d* levels into doubly degenerate $e_g$ level and triply degenerate $t_{2g}$ level. The predominant cause for such kind of splitting is symmetry of $t_{2g}$ orbitals (Ni $3d_{xy}$, $3d_{xz}$ and $3d_{yz}$) and $e_g$ orbitals (Ni $3d_{x^2-y^2}$ and $3d_{z^2}$). Existence of this symmetry suffers different interactions with O 2*p* orbitals which also subsequently splits the O 2*p* orbitals and into O $2p\pi$ and $2p\sigma$ bands [92]. $t_{2g}$ level lies below $e_g$ level and interaction between $e_g$ and oxygen $2p\sigma$ band is stronger than the same between $t_{2g}$ and O $2p\pi$ orbitals which results a repulsion between Ni 3d ($e_g$) and $2p\sigma$ orbital electrons [93]. It is reported by thakur *et al*. [94] that with increasing structural disorder or lattice distortion of $TiO_6$ octahedra the reconstruction of $t_{2g}$ and $e_g$ symmetry bands takes place. Overlap between *p* orbital of oxygen and *d* orbital takes place mainly due to lattice distortion [35]. From the aforementioned increasing GTF value (X-ray diffraction analysis), it is clear that lattice distortion is decreasing with increasing Ni doping. Therefore, lesser would be the lattice distortion/ structural disorder, lesser would be the orbital overlapping. So the impact of *p-d* hybridization will also get reduced which also reconfirms the presence of compressive strain in the local structure [79]. The gradual replacement of $Cd^{2+}$ with $Ni^{2+}$ with increasing Ni doping would certainly modify the unoccupancies of Ni 3*d* level by reducing the O 2*p* and Ni 3*d* hybridization. Therefore, due to reduced structural disorder, renovation of $t_{2g}$ and $e_g$ symmetry bands takes place in $NiO_6$ octahedra. In the simplest Kane k.p model, it is well suggested by Hui *et al.* that the band edge electron mass has a significant impact on *sp-d* hybridization. *P* (Momentum matrix element between conduction band and valence band) plays the main role in changing the *sp* Hamiltonian matrix [95]. *P* increases with decreasing band edge effective mass of the conduction electrons and the increment in *P* is a consequence of the diminishing *sp-d* hybridization [96]. However, it is redundant to say that the potential exchange between *s* and *d* orbital is not allowed by symmetry so the hybridization is mainly caused by *p-d* exchanged potential interaction. Therefore lowering of *p-d* hybridization lowers the band edge effective mass of electron and increases the electron-electron and electron-ionized impurity interaction described in (6) and (9) which subsequently increases the optical gap.

### 3.6 X-ray photoelectron spectroscopy:

Normalized XPS measurements have been performed for undoped, 5% Ni, 10% Ni, 40% Ni, 100% Ni doped CdO thin films with Al K$_\alpha$ monochromatic source. Collection of data has been done with a computer interfaced digital pulse counting circuit [97]. Calibration has been performed with C 1*s* peak set at 284.6 eV. Survey spectra for all thin films have been shown in Figure 9. Survey spectra are analysed using proper sensitivity factor for determination of actual stoichiometry. Peak fitting has been performed using CASA XPS software with Lorentzian (30%) and Gaussian (70%) for each component. Further, background removals are accomplished using Shirley type background for removal of the extrinsic loss structure [98]. From the survey scan spectra, it can be clearly seen that the binding energy peaks are getting sharper and prominent for Ni with increasing doping. As expected, no features of Cd orbital bands are found to observe in 100% Ni doped samples (i.e. pure NiO thin film). In Figure 10 (a) and 10 (d) stacked plots for Ni 2*p* and Cd 3*d* XPS spectra are shown and in Figure 10(b), XPS spectrum for NiO thin film has been shown separately for better understanding of the constituent multiplets and satellite peaks. The prime line structure of Ni 2*p* XPS spectra is being simulated with $Ni^0$, free ion multiplets i.e. $Ni^{2+}$ and $Ni^{3+}$ which has been well reported in the literature for NiO, Nickel hydroxide and oxyhydroxides [99–101]. In Figure 10(b) the binding energies (BEs) for $Ni^0$, $Ni^{2+}$ and $Ni^{3+}$ in Ni 2*p* XPS spectra are at 852.5, 854 and 856.4 eV respectively. A broad satellite peak is observed at 860.5 eV at Ni 2$p_{3/2}$ which is also well reported for NiO [102]. In literature this has been enunciated succinctly that main feature in Ni 2$p_{3/2}$ spectra i.e. at 854 eV for $Ni^{2+}$ is assigned to charge transfer phenomena in NiO. The broad satellite situated at 860.5 eV is based on the $cd^9L$ and unscreened $cd^8$ final state configuration (c is core hole, L is a ligand hole) [100,103]. From theoretical calculations, it has been observed that there are extra peaks generated for defects like $Ni^{3+}$ which we can observe clearly for 40% Ni and 100% Ni thin films XPS spectra unlike 5% and 10% Ni thin films. For 5% Ni thin film, we can observe a clear existence of $Ni^0$ and $Ni^{2+}$ states. The formation $Ni^0$ near surface region is well reported by weaver *et al*. with Ar ion sputtering [104]. Our experimental results over here are quite compatible with already reported theoretical results [102] utilizing multiple cluster calculation incorporating ligand charge transfer generates complex Ni 2p multilplet spectrum [105]. With increasing Ni doping the peak area for $Ni^0$, $Ni^{2+}$ and $Ni^{3+}$ in Ni 2p XPS spectra are found to be increasing and peaks and satellites features are shifting toward higher BE continuously. CdO exists in semiconducting phase while NiO is an insulating phase. Therefore, with increasing doping the phase is shifting from semiconducting to insulating which may generate a differential charging effect at the surface and lead to the shifting in BEs for peaks and satellite features [106]. This shifting is quite visible for O 1s and Cd 3d XPS spectra also. The values of the concerned peak position with peak area and FWHM are summarized at Table (S2) in ESM data sheet. It has been reported by Veenendaal and Swatzky that non-local screening effect in the strongly correlated transition metal oxides influences the XPS spectra and leads to the generation of satellite features [107,108]. As can be seen from Table (S2), satellites peak area is increasing with increasing Ni doping and gradually getting more and more prominent. In fact, such satellite features in Ni XPS spectra are often corroborated to excitation of a bound $3d$ electron to initially vacant states residing beyond Fermi level which is called shake-up states [109]. With increasing Ni doping, the prominent satellite peaks provides a signature that density of such kind of shake-up states is increasing, generating two correlates holes, (one is for core hole another is for $3d$ hole). Above Fermi level not only $3d$ unoccupied sub band states are there but also free-electron like $4sp$ states are present which hybridized itself with $3d$ character. With generation of core hole, $3d$ and $4sp$ hybridized states gradually shift below Fermi level due to core hole potential with increasing Ni doping and subsequently the probability of occupancy for those hybridized sates would be more and hybridization always appears to be more prominent for unoccupied states compared to the occupied ones [110]. Therefore, the increasing prominence of satellite features with increasing Ni doping provides an indirect signature of reduced orbital hybridization and compatible with experimental observation from XANES. Electronic properties investigation for transition metal oxide like NiO is always remained challenging due to its insulating behaviour, despite of its open $3d$ shell [111]. Mott-Hubbard theory has been incorporated for describing the insulating behaviour of the compounds like CoO, NiO, MnO etc. [92]. The electronic repulsion between localized, occupied donor type $d$ levels of Ni and unoccupied, extended conduction band states might be the possible reason for failure for traditional band properties and responsible for such kind of insulating nature in spite of having open shell system [112]. $3d$ transition metal compounds are categorised depending upon the Mott-Hubbard repulsion, $U$, charge transfer energy, $\Delta$, between ligand and metal, $3d$ orbital band width, ligand $2p$ band width etc. [113]. In NiO, this has been reported that it is a charge transfer insulator as $\Delta$ is lesser than $U$ in Zaanen-Sawatzky-Allen, ZSA, diagram[86]. NiO is having a significantly reduced charge transfer energy which means the energy essential to move a charge from ligand ion to metal cation is lesser. Therefore, charge transfer induced configurations in the XPS spectra get a significant contribution compared to coulomb interaction between $3d$ electrons in the outer shell and core hole at $2p$ shell [114].

Existence of plasmon losses in transition metal oxides due to excitation of consolidated oscillations of the $s$ and $p$ valence electrons is quite obvious [115]. For TM like MnO broad satellite feature at higher BE's is observed due to such kind of Plasmon losses. In present Ni $2p$ XPS spectrum for NiO apart from 860.5 eV satellite, two more satellite features have been observed at 873.3, 880.8 eV respectively. It has been reported that apart from shake–up states, aforementioned surface and bulk plasmon losses might be another reason behind such kind of satellite peaks. There are always some losses due to inter band and intra band transitions via relaxation of dipole selection rule in polycrystalline type samples with non-localized electrons [102]. However, Ni $2p$ XPS spectra is influenced by both surface as well as bulk with more prominent effect of bulk at the lower BE side [114]. Such kind of satellite feature also provides a signature of strong electron-electron correlation inside the system [116].

In Figure 10 (c) and (d) the XPS spectra for O $1s$ and Cd $3d$ are shown. In O $1s$ XPS spectra for 100% Ni thin film we observe a line at 529.7 eV which is not symmetric with a secondary broadening due to contaminants (mainly for moisture content in the air) at higher BE side. For undoped 4Cd thin films CdO, $CdO_2$ and $Cd(OH)_2$ phases are found to observe at BE of ~527.9, ~528.8, ~530.5 eV respectively. From Cd $3d$ XPS spectra also the existence of $CdO_2$ phase is quite visible. By continuous increasing doping, a gradual shift of O $1s$ signal between two distinct phases i.e. from CdO to NiO can be observed clearly and the contamination peak also exists for both of the phases.

Even after this whole explanation of XPS fitting spectra for NiO, unambiguous deconvolution of spectral features is quite difficult for Koster-Kronig Augur decay regarding the transition metal hole state[114]. However, there might be impact of inter band transition between unoccupied $d$ states and partially filled valence band $d$ states which might make the deconvolution much more complicated [117].

### 3.7 Theoretical Calculation:

The fully optimized structures of $Ni_xCd_{1-x}O$ are depicted in Figure 11. The optimized lattice parameter of CdO unit cell is a=b=c=4.78 Å and bond length between Cd-O is 2.38 Å which matches well with previously reported work [118]. While 100 % Ni doping in place of Cd in rock-salt structure of CdO change and reduces the lattice parameter a=b=c=4.16 Å and Ni-O bond length is 2.08 Å and these parameters are in good agreement with previous work[30]. Figure 11(a) shows the optimized structure of rock-salt CdO and 5% Ni doping in CdO, the structure is slightly distorted, changed their bond lengths and lattice parameter which is presented in Figure 11 (b). While 100 % doping of Ni in place of Cd in rock-salt structure of CdO, it reduces 12.97 % lattice parameter and it shown in Figure 11(c). The results are in complete conformity with X-ray diffraction analysis.

Furthermore, to understand the effect of Ni doping in place of Cd in rock-salt CdO, we have computed the electronic band structures and projected density of states (PDOS) for $Ni_xCd_{1-x}O$ with increasing doping percentage, as shown in figure 12-14. The rock-salt $Ni_xCd_{1-x}O$ (x=0.0) shows the semiconducting behavior with 0.99 eV band gap from M to Γ-point (indirect band gap) while the direct band gap is 1.14 eV at Γ-point as presented in Figure 2(a). To understand the orbital contribution in the electronic band structure, we have calculated the projected density of state and we can clearly observe that the peaks corresponding to the Cd $4d$, O $2s$ and O $2p$ states as shown in Figure 12(a). The splitting of states of rock-salt $Ni_xCd_{1-x}O$ (for x=0.0) such as Cd $4d$ and O $2s$ states are shifted at deep energy levels in the valence band and get broadened too. There is notable orbital mixing between Cd $4d$ and O $2p$ states appear near the Fermi level. Sharp peaks of O $2p$ states are found near the Fermi level in VBM. Orbital hybridization between Cd 4$d$ and O 2$p$ is succinctly observed in Figure 12(a). However, O 2$p$ states are significantly dominating in this orbital mixing near Fermi level.

Now, we have discussed the effect of doping in rock-salt CdO structure. In Figure 12(b) shows the 5% doped one and corresponding band structure is semiconducting nature and have 0.59 eV indirect band gap from L to Γ-point. The Ni 3$d$ states appeared to be near -7 eV at the VBM. Even this lower doping percentage has a significant impact upon O 2$p$ orbital states which get broadened. However, at 5% Ni doping, there is not significant orbital mixing between Ni 3$d$ and O 2$p$ states is observed. For 10% Ni doping, the structures of Ni 3$d$ states reflect hybridization with Cd 4$d$ and O 2$p$ states in the VBM to some extent. Furthermore, Ni 3$d$ states are shifted towards the Fermi levels which can be observed at Figure 12(c).

In Figure 13 (a), (b) shows the electronic band structure and projected density of states where Ni concentration is x=0.2 and 0.4, which have strongly affected the electronic properties of rock-salt CdO structure. Now, with improved Ni percentage in Ni$_x$Cd$_{1-x}$O, splitting of Ni 3$d$ electronic states are observed which get prominent and broadened with increasing doping, up to 80%. However, for 20% and 40%, the orbital overlapping between Ni 3$d$ and O 2$p$ states gets reduced significantly and simultaneously the mixing between Cd 4$d$ and O 2$p$ also deteriorates with notable reduction of density of states for Cd 4$d$ electronic states at VBM. Furthermore, at 80 % of Ni concentration then the Ni 3$d$ states appear to be completely separated from O 2$p$ electronic states and reflect high intensity around -7 eV in VBM and also enhanced the band gap which is found 0.89 eV indirect band gap from L to Γ-point and direct band gap is 1.04 eV at Γ point as shown in Figure 14(a). From the PDOS of 80% concentration of Ni, a strong appearance of Ni 3$d$ states at conduction band maximum (CBM) is also observed which reflects robust interaction between those localized $d$ states and extended conduction band states in Figure 14(a). The gradual separation of Ni 3$d$ electronic states from O 2$p$ states delivers a direct evidence of reduced orbital hybridization which is in complete agreement with all experimental outcomes and results a band gap renormalization.

After the 100 % Ni concentration in rock-salt Ni$_x$Cd$_{1-x}$O, the O 2$s$ states appear at very deep energy level in VBM at Figure 14(b). The structures of Ni 3$d$ states are heavily hybridized with O 2$p$ states but splitting of O 2$s$ states shifted down in energy. We can clearly observe that the peaks correspond to Ni 3$d$ and O 2$p$ states around the Fermi level. The analysis of core level spectroscopy reflects the reduction of orbital hybridization up to 100% Ni doping. But PDOS reflects strong $p$-$d$ overlapping for 100% Ni thin film. As NiO also exists in rocksalt structure with octahedral symmetry, therefore, again it also shows a strong $p$-$d$ hybridization like rocksalt CdO. However, the trend of orbital mixing, density of states and band structure is completely different with the rocksalt structure of pure CdO. During the increments of Ni percentage in rock-salt $Ni_xCd_{1-x}O$, the composite system changes it's Fermi level position with respect to the vacuum level and electronic band gap as shown in Figure 15.

It is reported by Francis *et al*. [37] that after 40% of Ni doping the composite system becomes completely insulating. However, from the existing band structure calculations an strong hybridization of at valence band between oxygen 2*p* and partially filled *d* states is claimed which anticipates a resulting p-type conduction [119,120]. The same has been observed in the projected density of states for NiO in Figure 13(c) in our calculation. In Figure 14, we observe the change of Fermi energy with respect to the vacuum in increasing with enhanced doping percentage. However, there is an anomalous behavior at 20% Ni doping where the change takes place in opposite direction. This change of Fermi energy position can be explained by incorporating the concept of ADM [21]. According to this model, point defects, dopants are classified in three domains, i.e. delocalized, localized and shallow defects. The corresponding energy levels of the concerned dopants or defects are dependent upon band edge energies which can be changed via external perturbation like modifying composition or exerting hydrostatic pressure. The transitional type impurities come into the class of localized defects with localized *d* shells. The energy levels of those localized *d* shell electrons are insensitive to band edge energies and that's why these energy levels are considered as reference levels for determining band offset phenomena in III-IV and II-VI compounds [121]. For CdO, $E_{FS}$ is situated at 4.9 eV below the vacuum level which is near about 1 eV above the CBM and which is favorable condition for n-type conductivity [37]. As a linear dependence of band edge with composition CBM is expected to cross $E_{FS}$ at 20% of doping and tendency of n-type conductivity reduces significantly. Theoretically, it is anticipated that p-type conductivity is possible in NiO which has been explained by different models like Mott-Hubbard insulator [122] and charge-transfer insulator[86]. Theoretically, the Ni vacancies which are having energy levels at the vicinity to the VBM are the prime reason behind such *p*-type character [123]. However, experimentally it is quite difficult to achieve *p*-type conductivity by doping in materials like CdS, CdSe etc. In present case, we can observe that Fermi level position is changing from 0 to 5.5 eV which provides a signature that with Ni doping the conductivity is changing to *p*-type from *n*-type. Therefore, crossing $E_{FS}$, Fermi level position goes to the vicinity of VBM. From this observation, one can assert that 5.5 eV is the allowed band of Fermi energy. At 20% of Ni doping CBM crosses the $E_{FS}$ and subsequently mobility and electron concentration reduces significantly. Always intrinsic tendency for Fermi level of any system to follow $E_{FS}$ to minimize the overall entropy of the system. This might be the possible reason for anomalous change of Fermi level position at that concerned compositional juncture. In Figure 14, up to 10% doping we observe that band gap is reducing and 20% onwards both direct and indirect band gap keeps on increasing. It might be possible that due to surface electron accumulation phenomena for CdO which has not been considered during density of state calculation has significant impact in band gap renormalization via B.M. shift effect. However, with Ni doping electron concentration reduces and electron-electron, electron-ionized impurity interaction comes into the picture for explaining experimental observation. But as we observe band gap reduction in theoretical calculation until CBM crosses $E_{FS}$, therefore we can anticipate that lower doping percentage the aforementioned interaction doesn't influence the calculation significantly.

**Conclusion:**

In conclusion, we report the optical bandgap widening in $Ni_xCd_{1-x}O$ thin films with increasing Ni concentration. In particular, cationic substitution by replacement of $Cd^{2+}$ with $Ni^{2+}$ ion leads to alter the local coordination environment of the CdO matrix by development of local compressive strain with enhancement in electron-electron, electron-impurity and electron-phonon scattering. In fact, lowering the degree of *p-d* hybridization between O *p* and Cd/Ni *d* orbitals with increasing Ni doping has been elucidated by the detail analysis of O *K*-edge XANES spectra and XPS Ni 2*p* spectra. The decrease in electron mass at conduction band edge and *p-d* hybridization states lead to modify the distribution of Fermi level energy driven electron concentration which subsequently affects the electron-electron, electron-ionized impurity interaction and widens the optical gap in $Ni_xCd_{1-x}O$ thin films. PDOS calculations reflects the orbital separation between O 2*p* and Ni/Cd *d* orbitals which is in complete agreement with our experimental observation.

**ACKNOWLEDGMENTS**

Authors are grateful to the Director (IUAC), Director (UGC-CSR Indore) for their encouragement and moral support for extending experimental facilities including the synchrotron facility at RRCAT Indore. Authors are also thankful to Dr. Fouran Singh for discussion, Dr. Indra Sulania for experimental support in AFM investigations, Mr. G. R. Umapathy for RBS measurements. Authors are grateful to Dr. Dinesh Kumar Shukla for oxygen K edge measurements at BL01 in Raja Ramnna Centre for advanced rechonology (RRCAT). One of the authors (A. Das) acknowledges to Senior Research Fellowship (SRF) Grant Number-F.2-91/1998(SA-1) from University Grant Commission, New Delhi. D. Singh and R. Ahuja would like to thanks the Carl Tryggers Stiftelse for Vetenskaplig Forskning (CTS) & Swedish Research Council (VR) for financial support. SNIC and SNAC are acknowledged for providing the computing facilities. DST, Govt. Of India is also gratefully acknowledged for providing FE-SEM through nano-mission project and for granting Science and Engineering Research Board (SERB) project (SB/EMEQ-122/2013).

**TABLE AND FIGURE CAPTIONS:**

**Table 1:** Numerical values of different parameters for pure (4Cd) and Ni doped CdO thin films.

**Figure 1:** Recorded Rutherford backscattering spectra (black curve) and corresponding simulated spectra (red curve) of (a) 4Cd, (b) 40% Ni, (c) 80% Ni and (d) 100% Ni thin films, respectively.

**Figure 2:** Typical x-ray diffraction spectra of pure CdO (4Cd) and Ni doped CdO thin films.

**Figure 3:** Plot for Goldschimdt's Tolerance Factor with increasing Ni doping percentage.

**Figure 4:** Atomic force microscopy images of (a) 4Cd, (b) 5% Ni, (c) 20% Ni and (d) 80% Ni thin films, respectively.

**Figure 5:** (a) Transmittance spectra and corresponding (b) Tauc plot of 4Cd, 3% Ni, 5% Ni, 10%Ni, 20% Ni, 40% Ni, 80% Ni and 100% Ni thin films, respectively, showing the variation in optical band gap with increasing Ni doping concentration. Here the optical band gap is determined through linear extrapolation in Tauc plot.

**Figure 6:** Bowing equation fitting, showing the variation in optical band gap with increasing Ni concentration.

**Figure 7:** Raman spectra of pure CdO and Ni doped CdO ($Ni_xCd_{1-x}O$; x= 0, 0.03, 0.05, 0.1, 0.2, 0.4, 0.8, 1) thin films, annealed at 400 °C.

**Figure 8:** Soft X-ray absorption spectra for the O *K* edges for 4Cd, 5% Ni, 10% Ni, 80% Ni, 100% Ni thin films

**Figure 9:** XPS survey scan spectra acquired for 4Cd, 5% Ni, 10% Ni, 40% Ni, 100% Ni thin films.

**Figure 10:** Typical-high resolution XPS spectra of pure and Ni doped CdO thin films recorded near (a) Ni-*2p* region, (c) Cd-*3d* region, and (d) O-*1s* region, respectively. For better clarity, magnified view of Ni-*2p* spectra for 100% Ni doped CdO thin film is also shown in (b).

**Figure 11:** The optimized structure of rock-salt structure of (a) CdO, (b) 5 % Ni doping in CdO and (c) 100 % Ni doping in CdO.

**Figure 12:** The electronic band structure and corresponding projected density of states (PDOS) of rock-salt structure of (a) pure CdO, (b) 5 % Ni doping in place of Cd in CdO and (c) 10 % Ni doping in place of Cd in CdO.

**Figure 13:** The electronic band structure and corresponding projected density of states (PDOS) of rock-salt structure of (a) 20 % Ni doping in place of Cd in CdO and (b) 40 % Ni doping in place of Cd in CdO.

**Figure 14:** The electronic band structure and corresponding projected density of states (PDOS) of rock-salt structure of (a) 80 % Ni doping in place of Cd in CdO and (b) 100 % Ni doping in place of Cd in CdO.

**Figure 15:** The variation of Fermi energy, direct and indirect band gap of $Ni_xCd_{1-x}O$ where x varies from 0 to 1 (x=0.0, 0.05, 0.1, 0.2, 0.4, 0.8, 1.0). The black, red and blue lines display the variation of Fermi energy, direct band gap and indirect band gap, respectively.

**Table 1**

| Sample name | Particle diameter (nm) | Band gap (eV) | *rms* roughness (nm) | Thickness (nm) | GTF factor |
|---|---|---|---|---|---|
| 4Cd | 19 | 2.42 | 2.7 | 180 | 0.707 |
| 3% Ni | 13.8 | 2.52 | *n.m.* | 170 | 0.710 |
| 5% Ni | 13.2 | 2.57 | 1.9 | 164 | 0.710 |
| 10% Ni | 9.4 | 2.59 | 0.88 | 167 | 0.716 |
| 20% Ni | 9.4 | 2.60 | 0.87 | 155 | 0.729 |
| 40% Ni | 5.9 | 2.69 | 1.39 | 140 | 0.752 |
| 80% Ni | *n.a.* | 3.65 | *n.m.* | 150 | 0.803 |
| 100% Ni | *n.a.* | 3.70 | *n.m.* | 175 | 0.832 |

*n.a. = not applicable*

*n.m. = not measured*

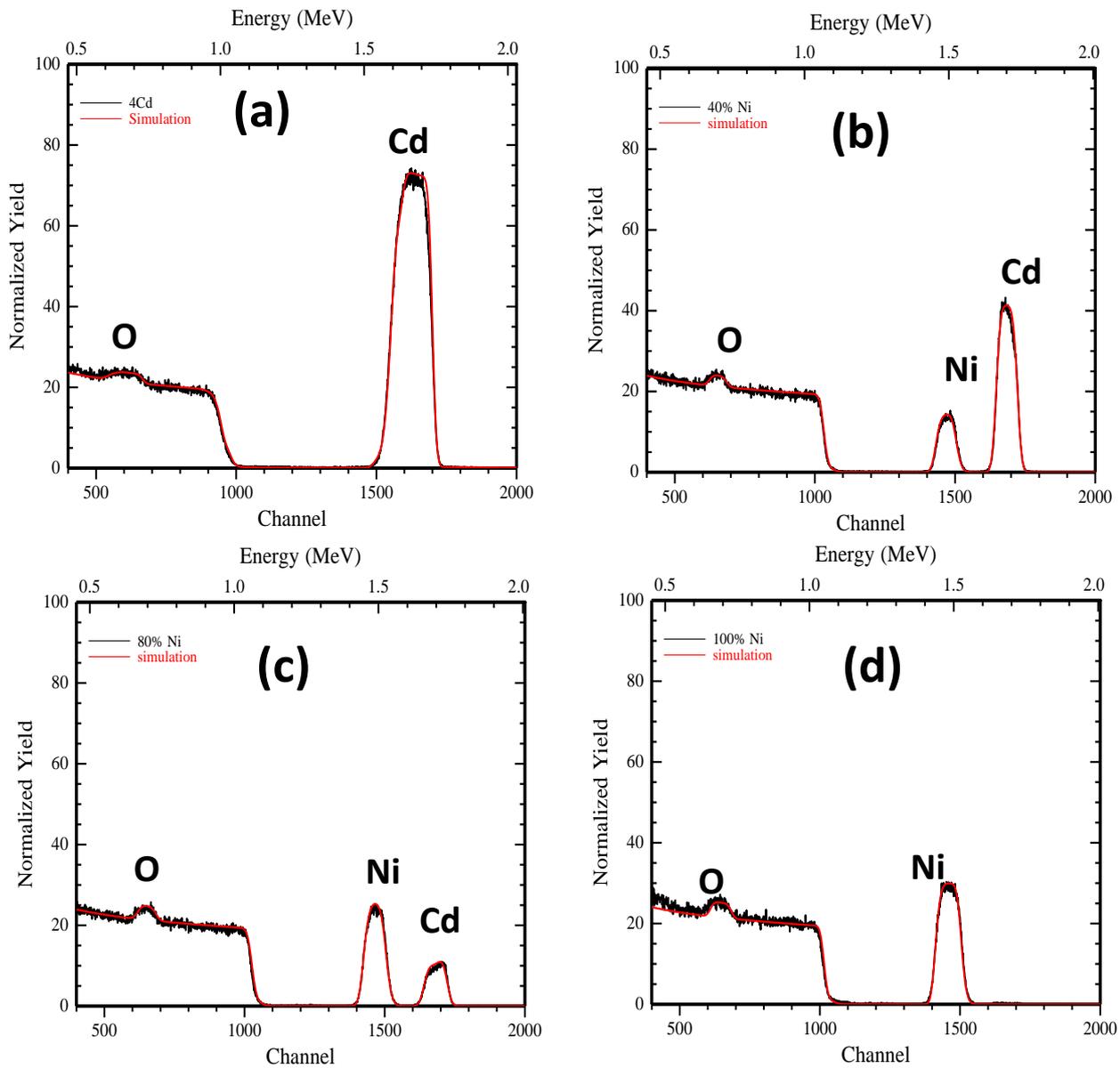

**Figure-1**

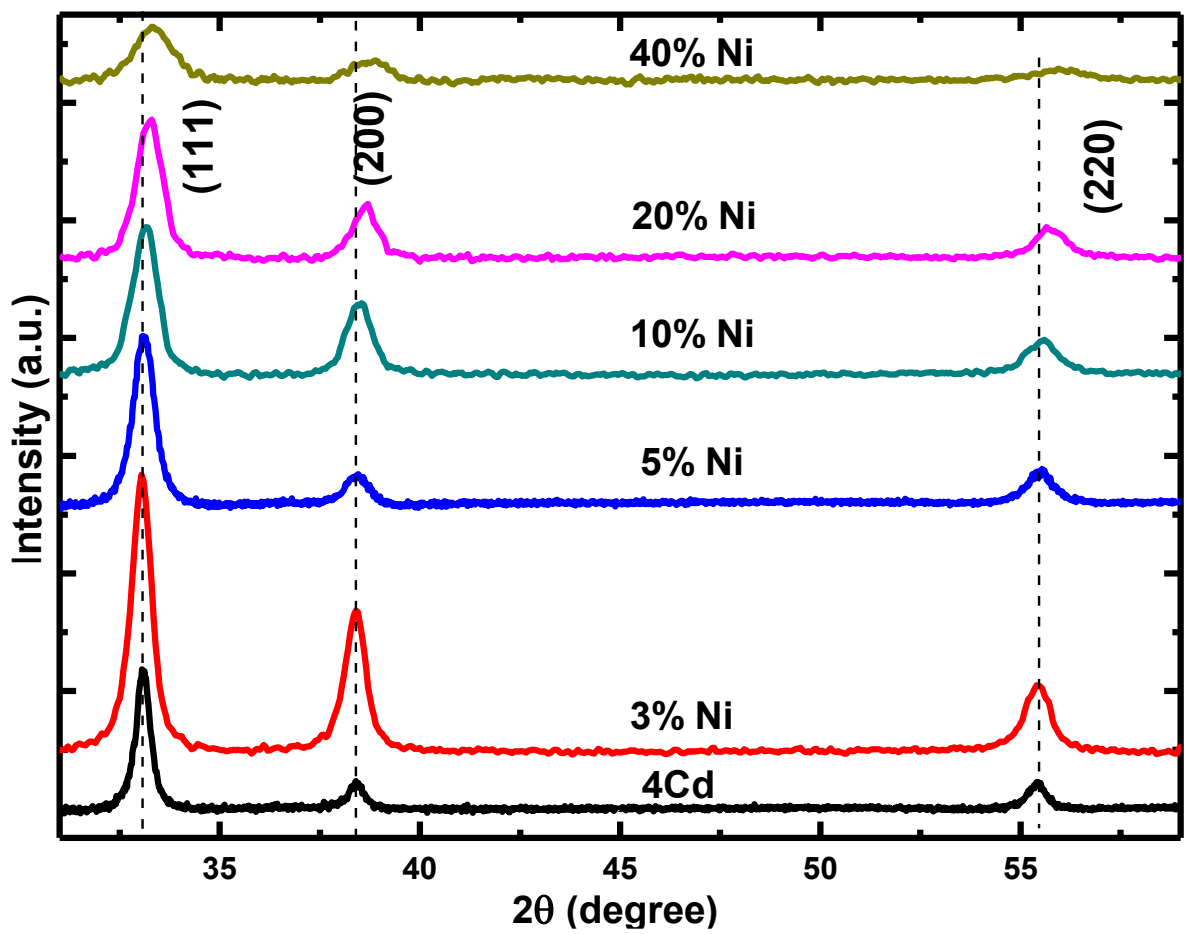

Figure-2

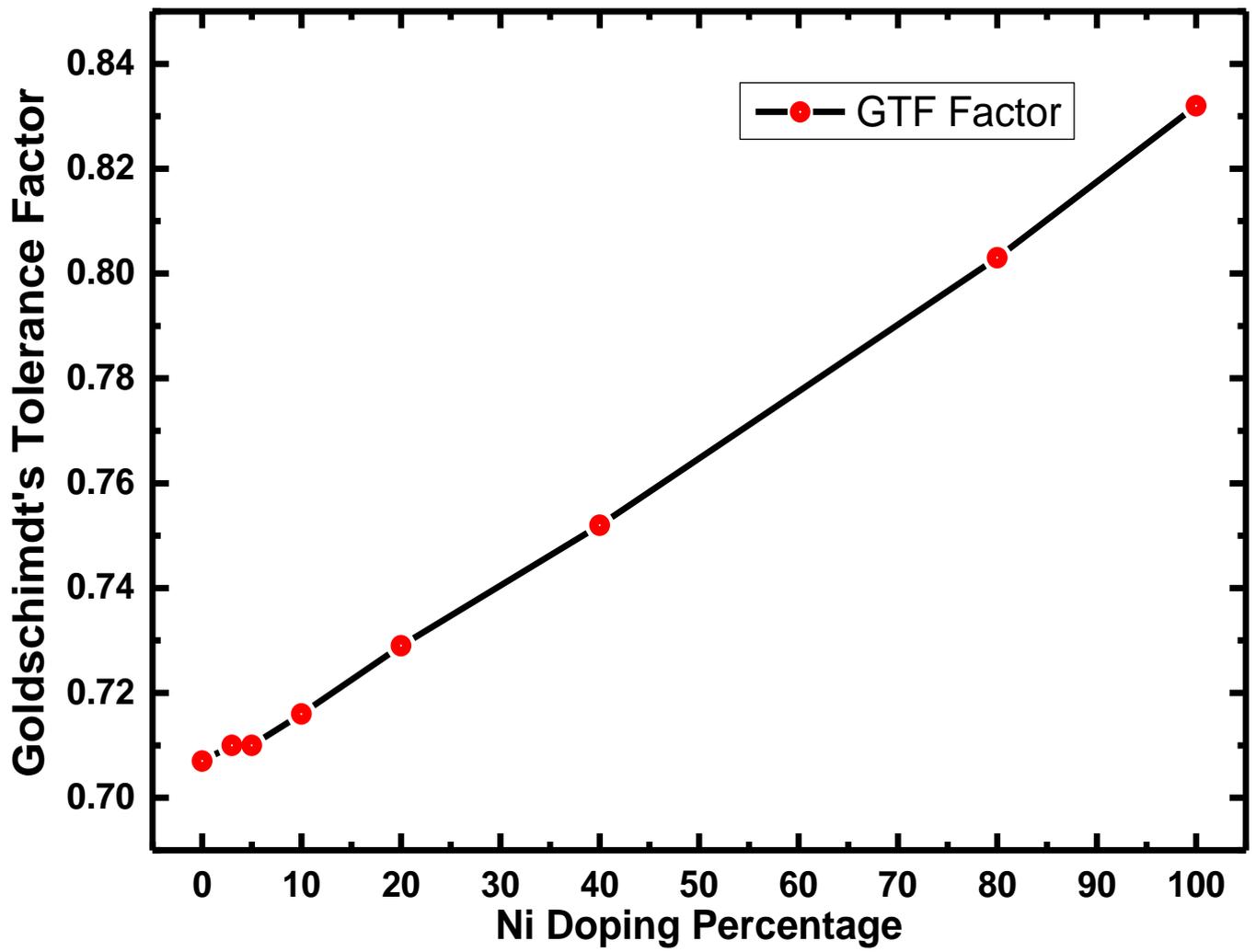

**Figure-3**

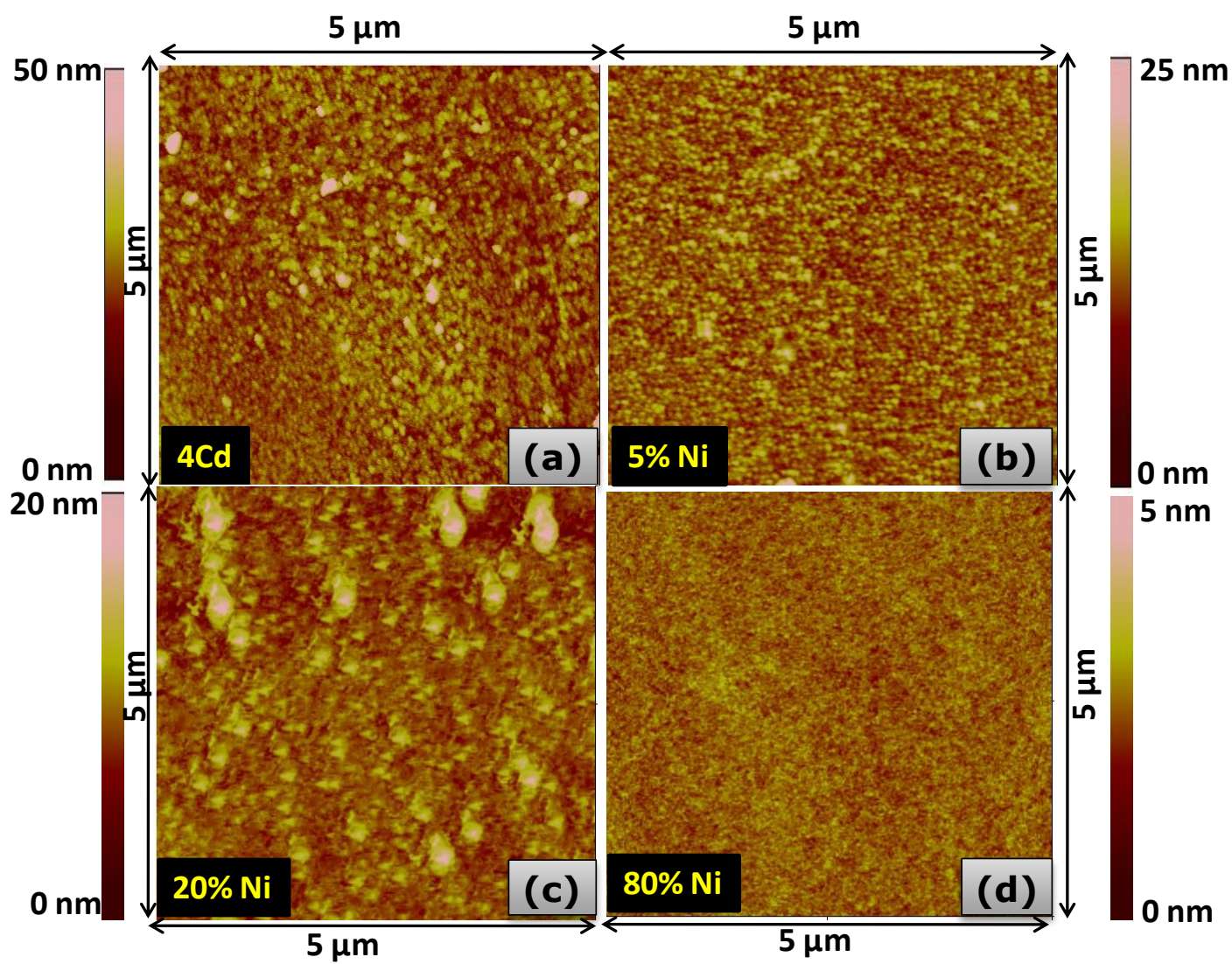

**Figure-4**

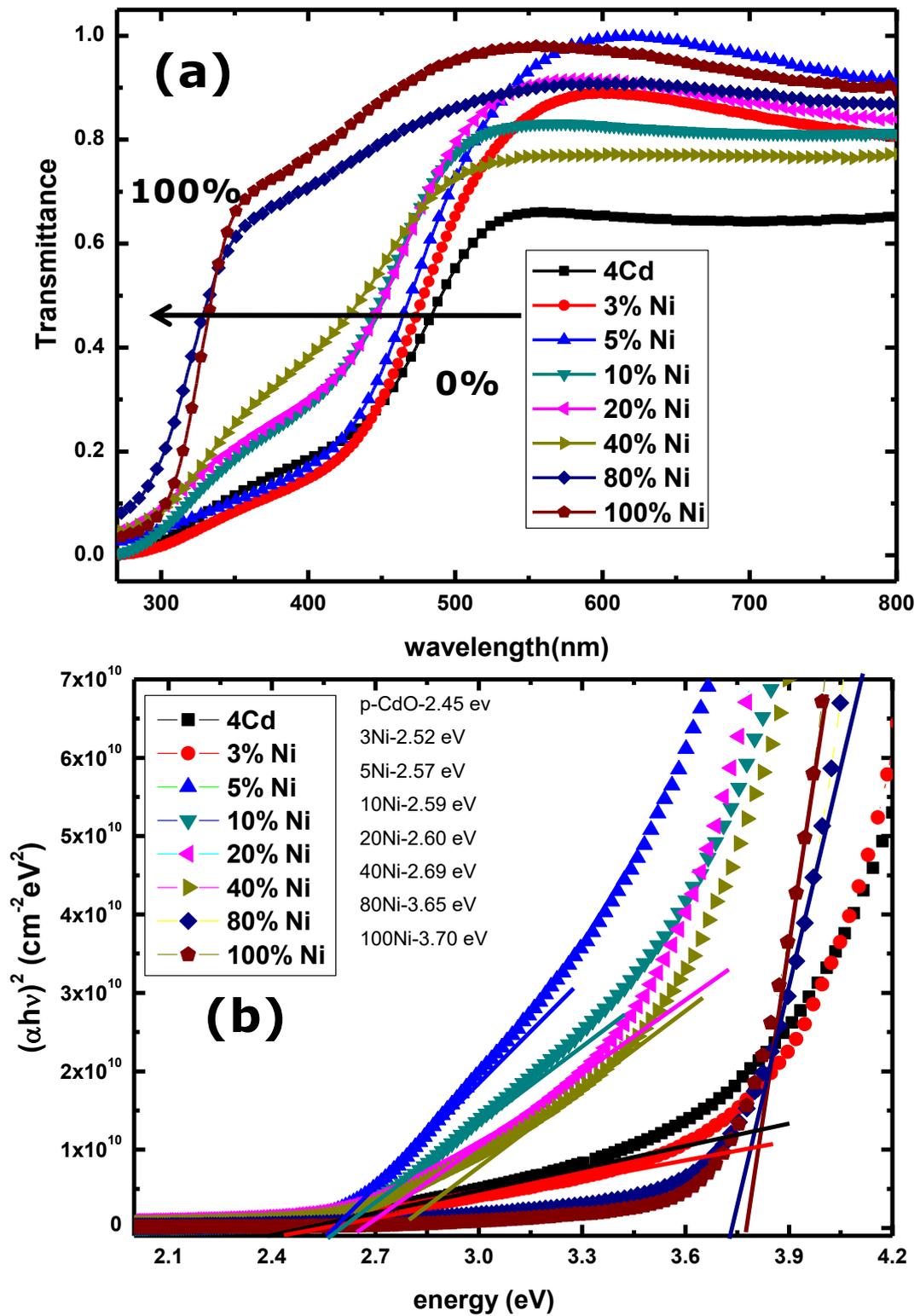

Figure-5

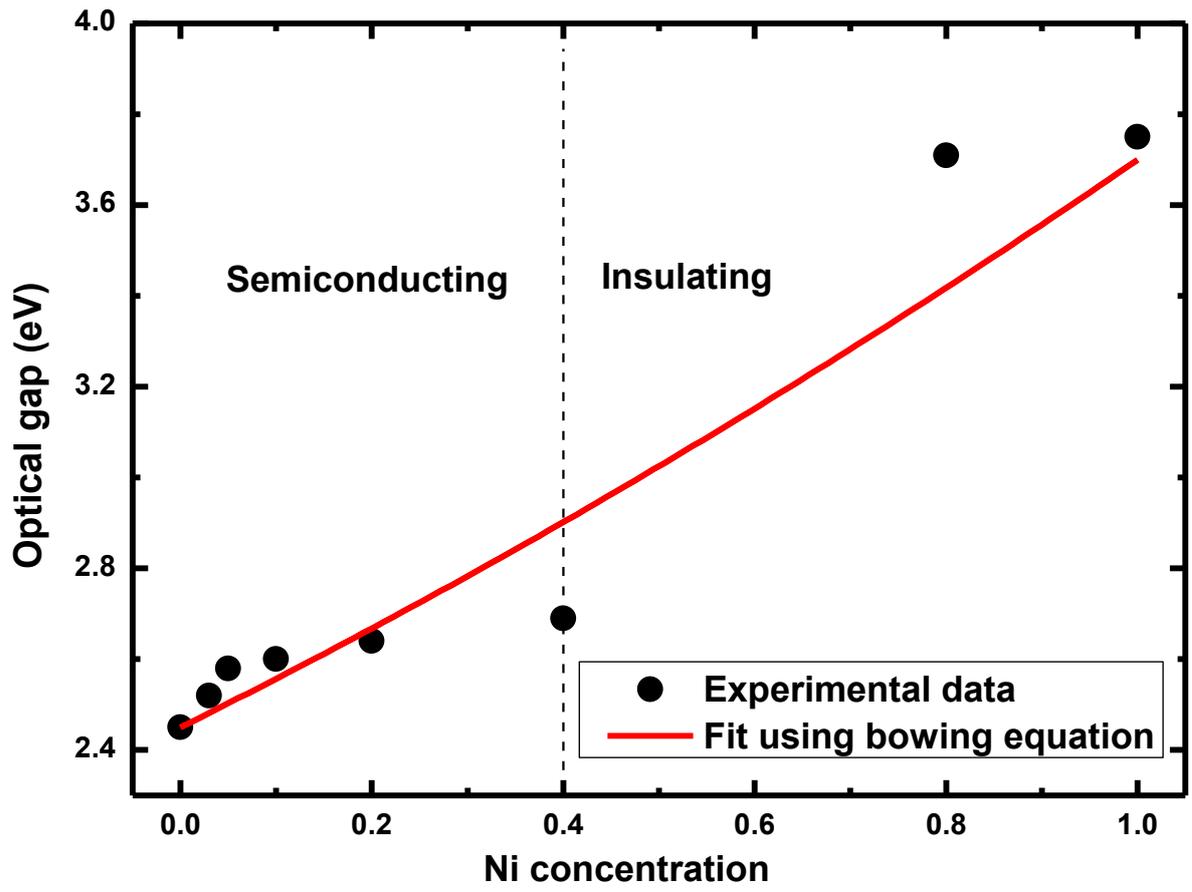

**Figure-6**

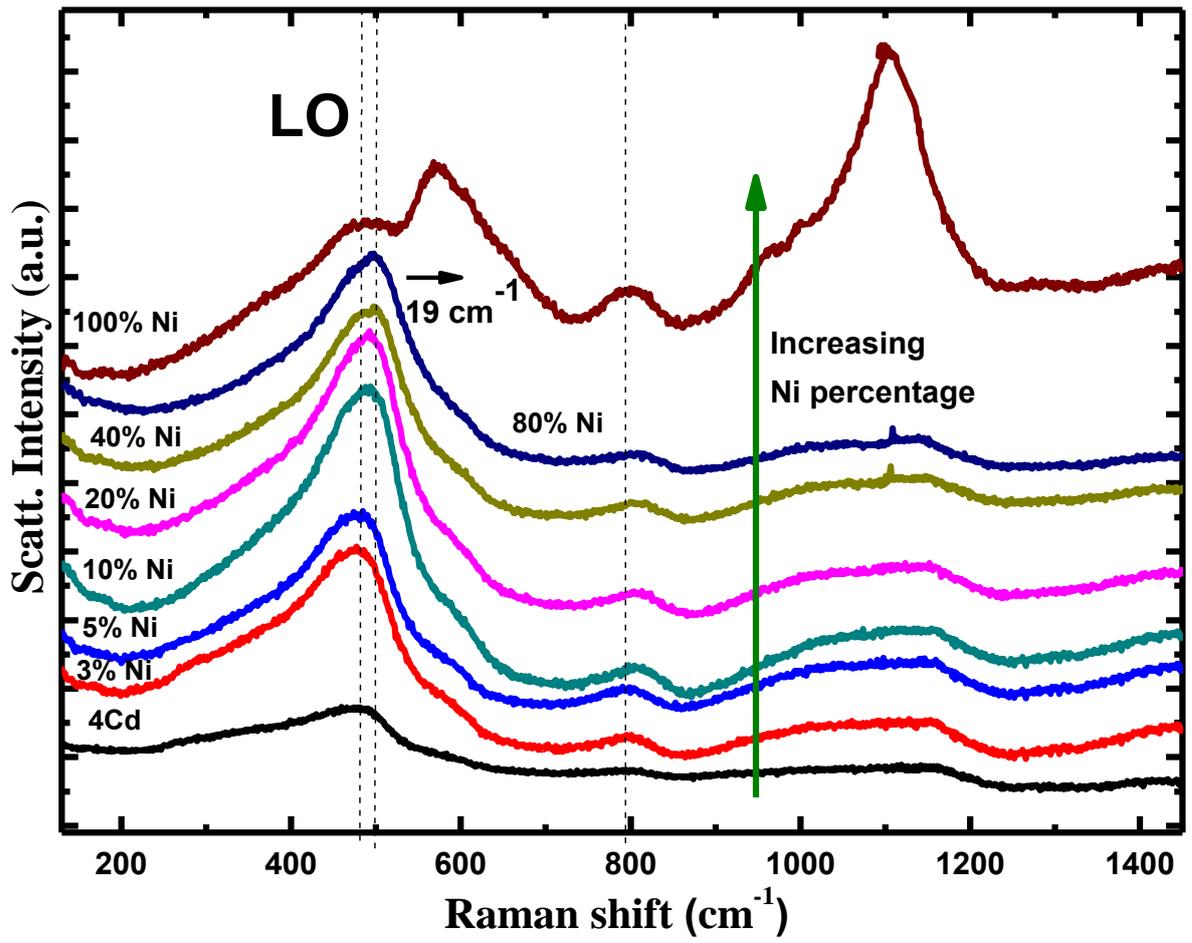

Figure-7

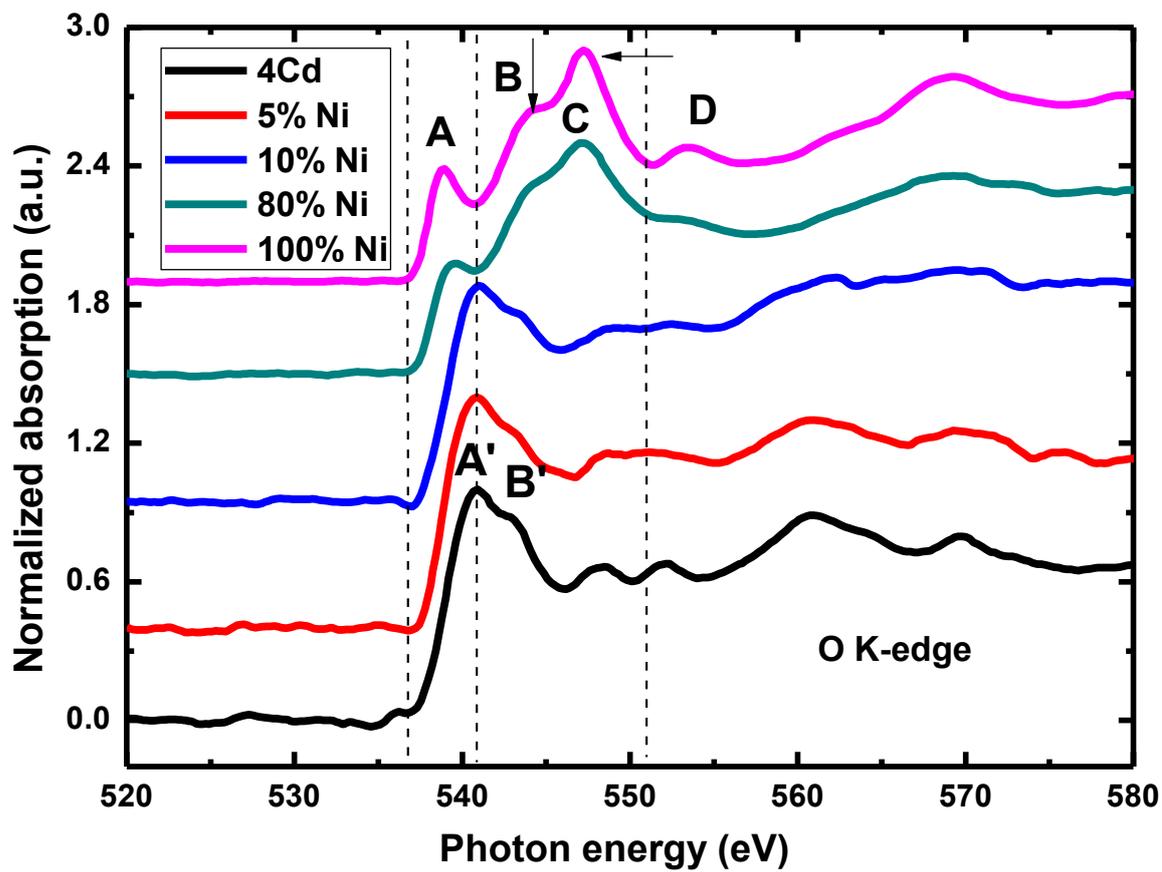

**Figure-8**

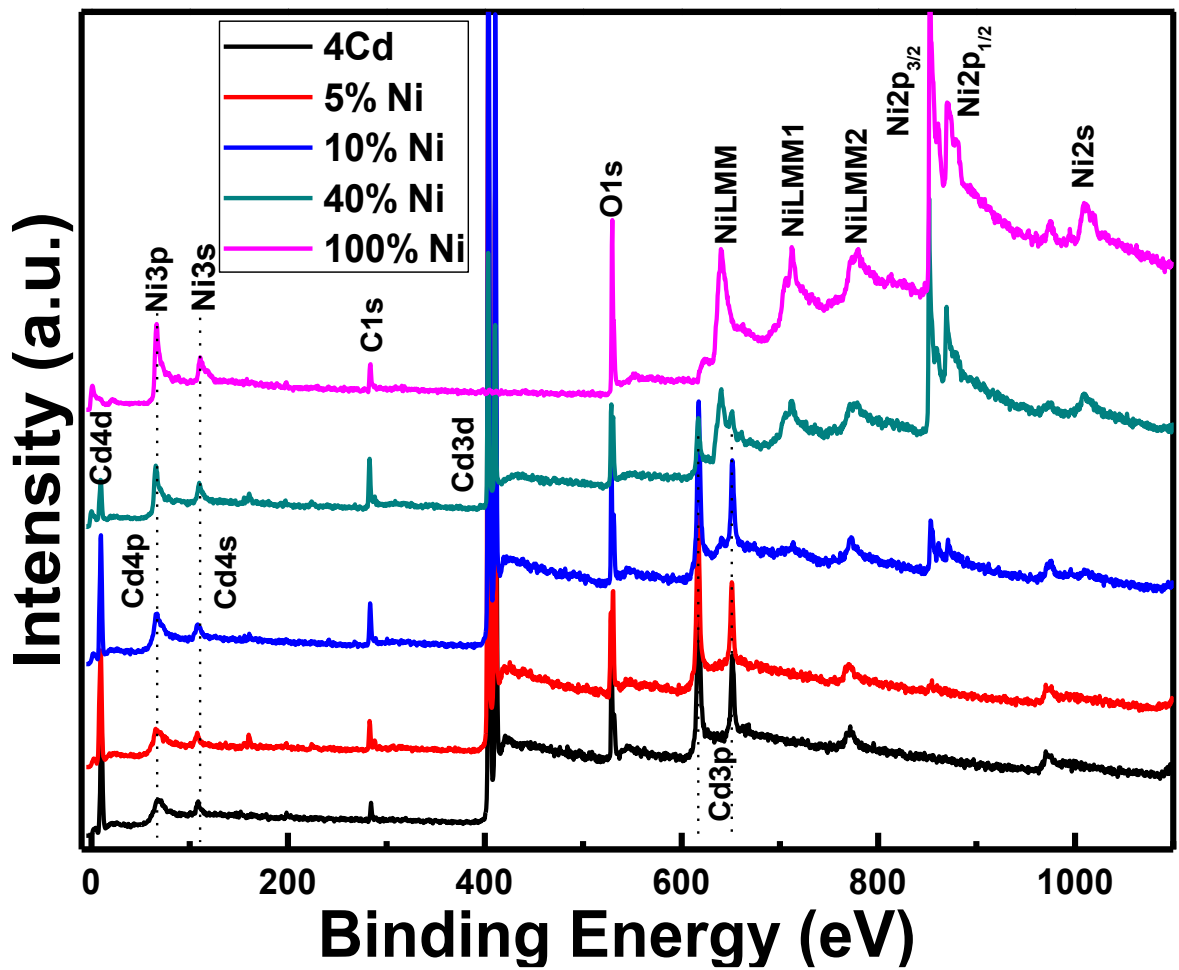

**Figure-9**

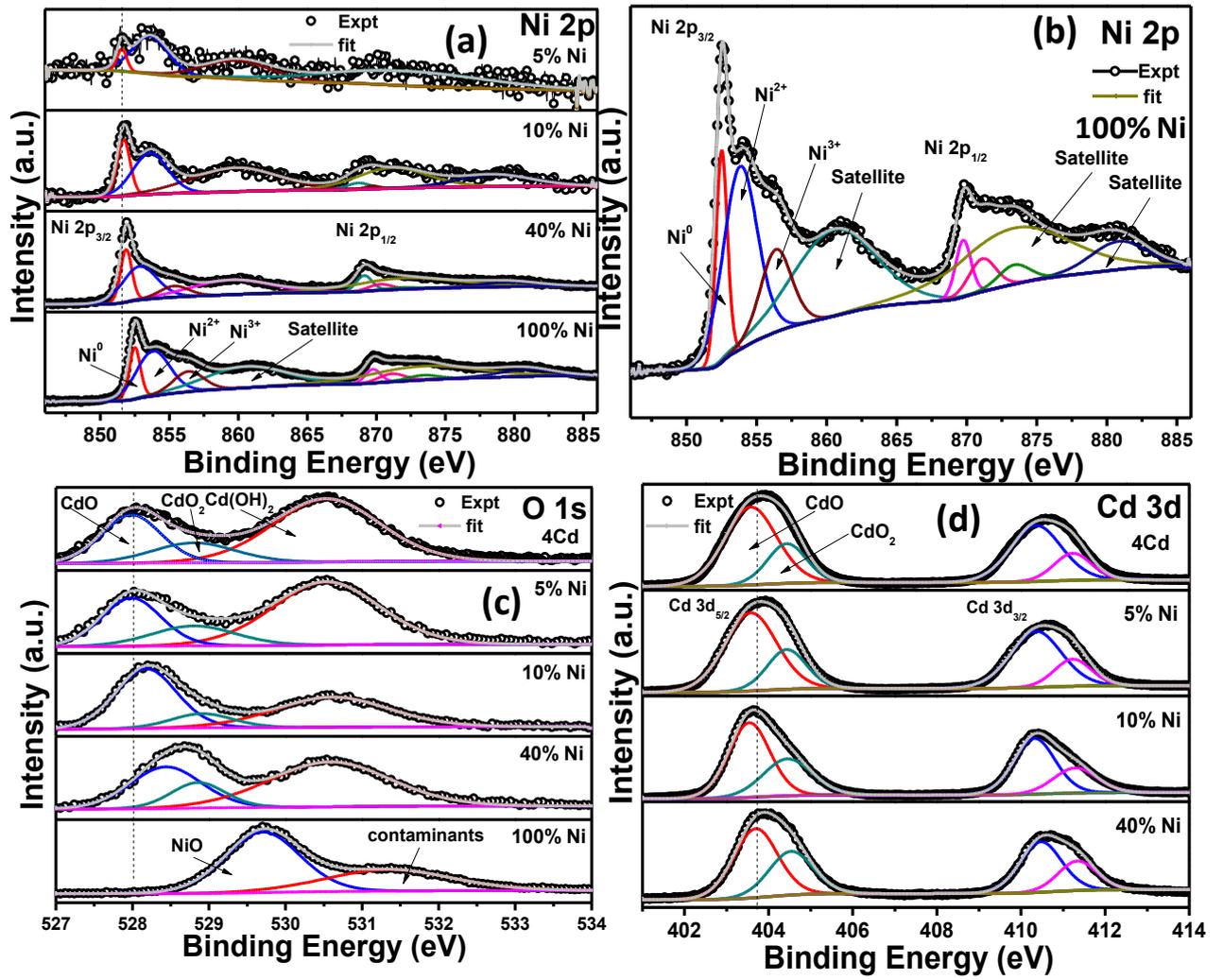

**Figure-10**

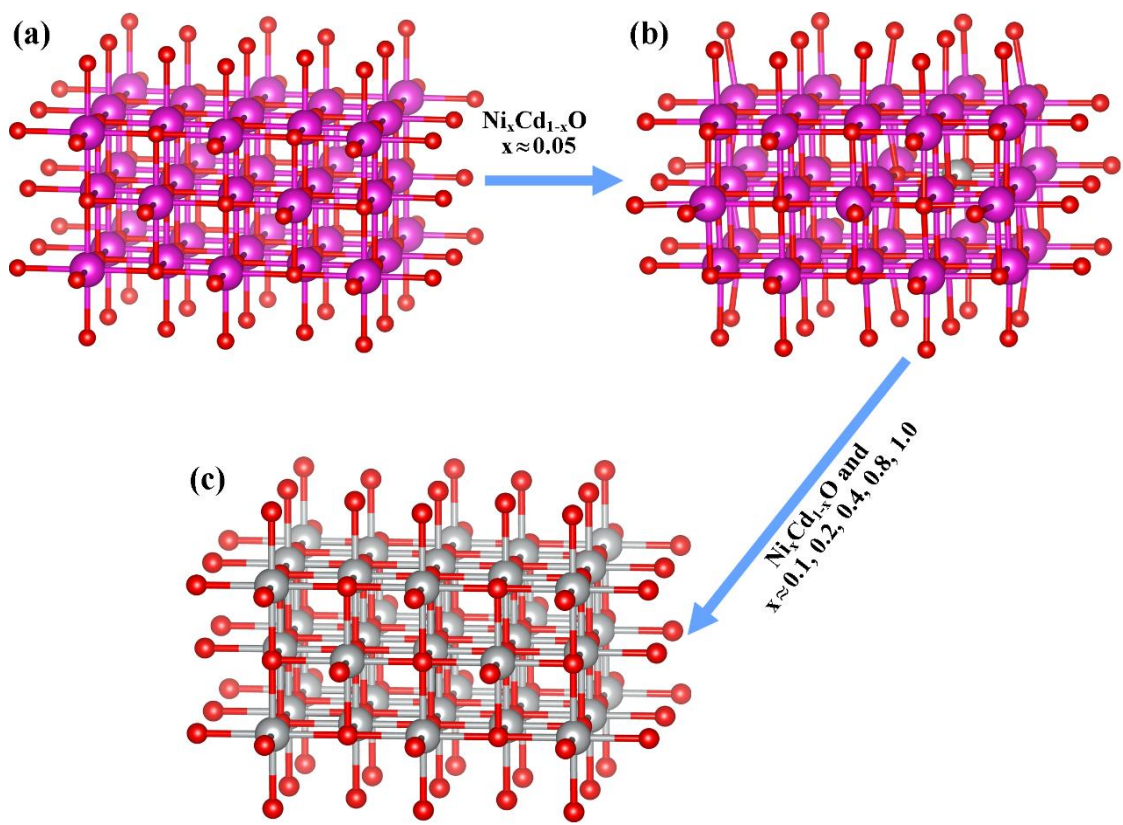

**Figure-11**

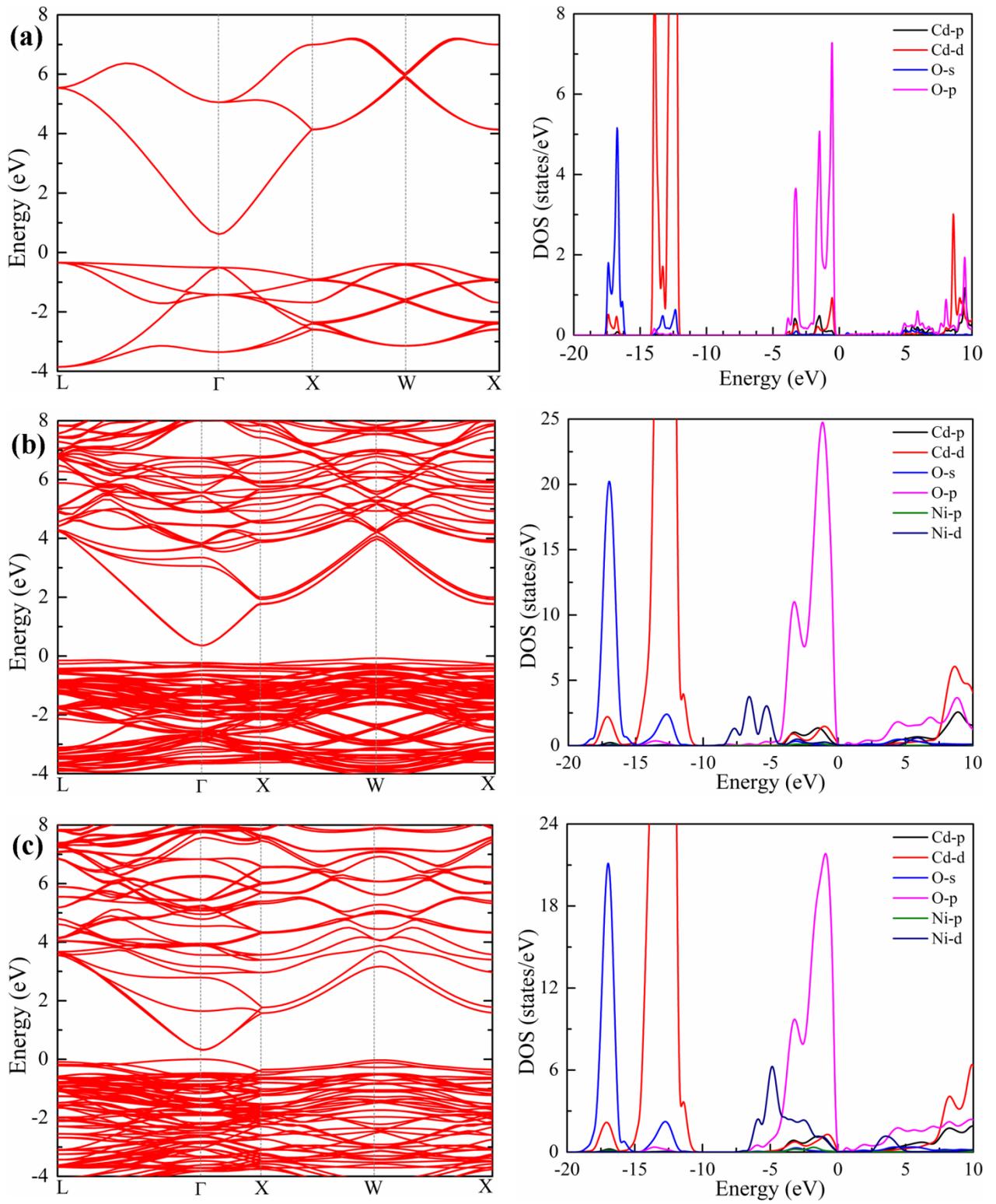

**Figure-12**

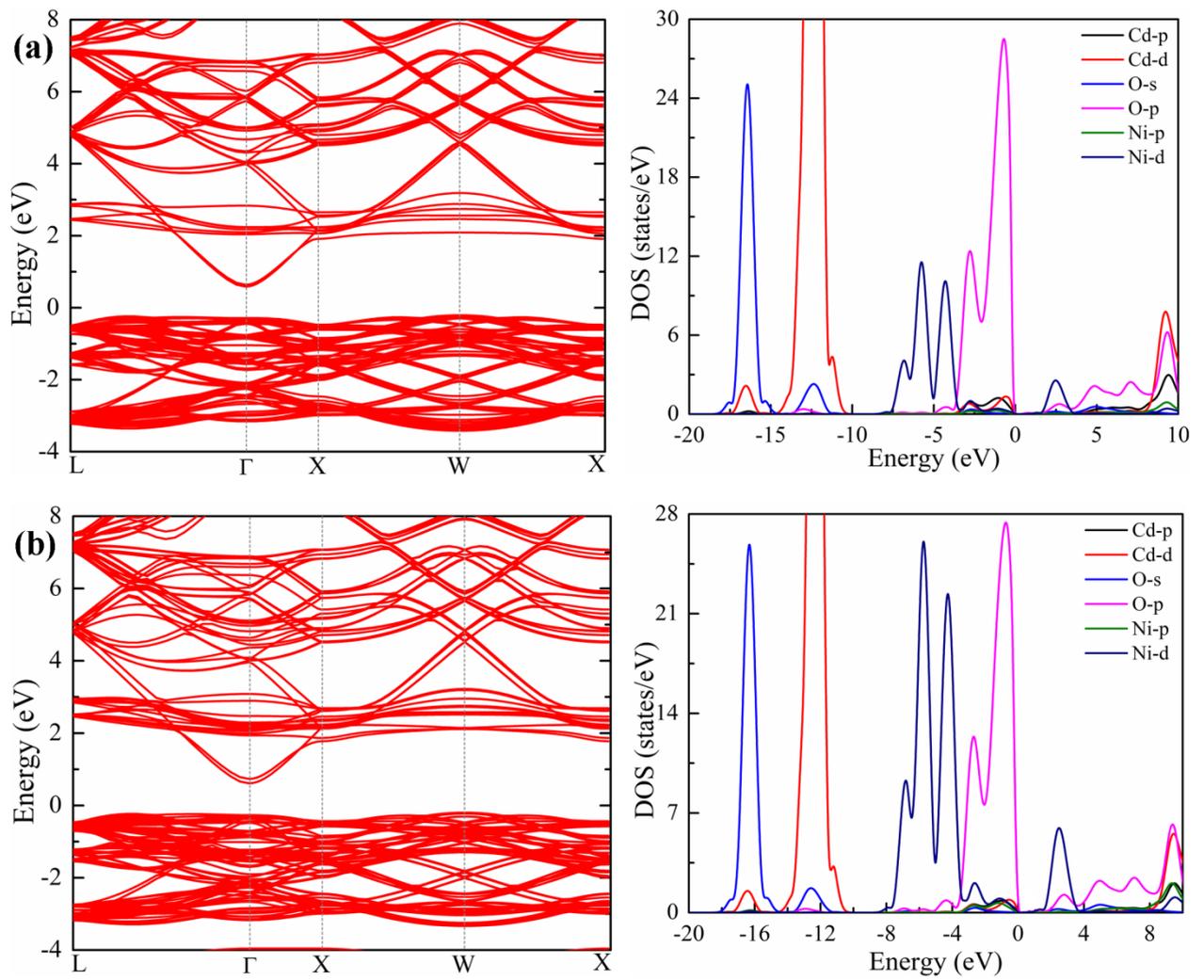

**Figure-13**

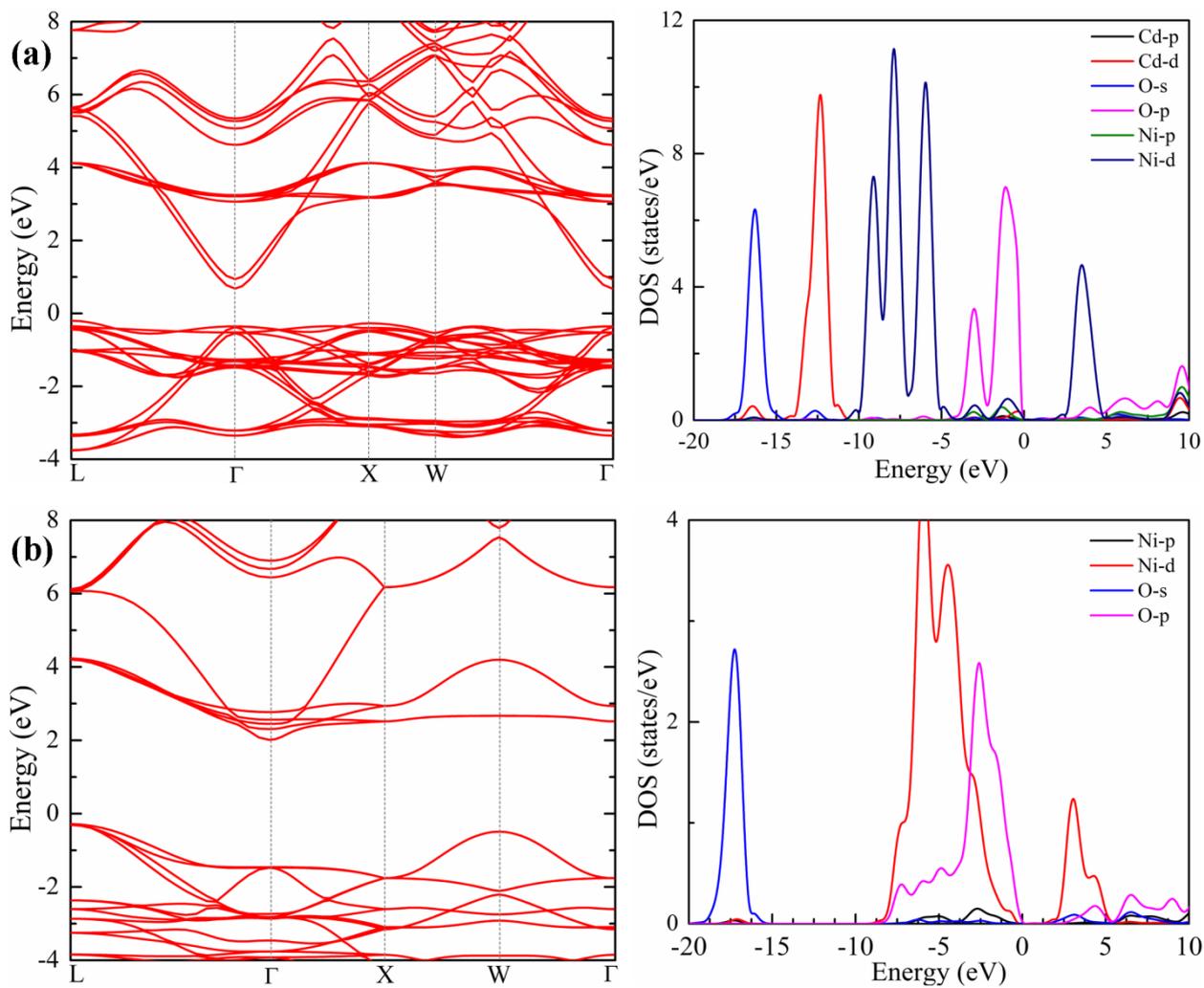

**Figure-14**

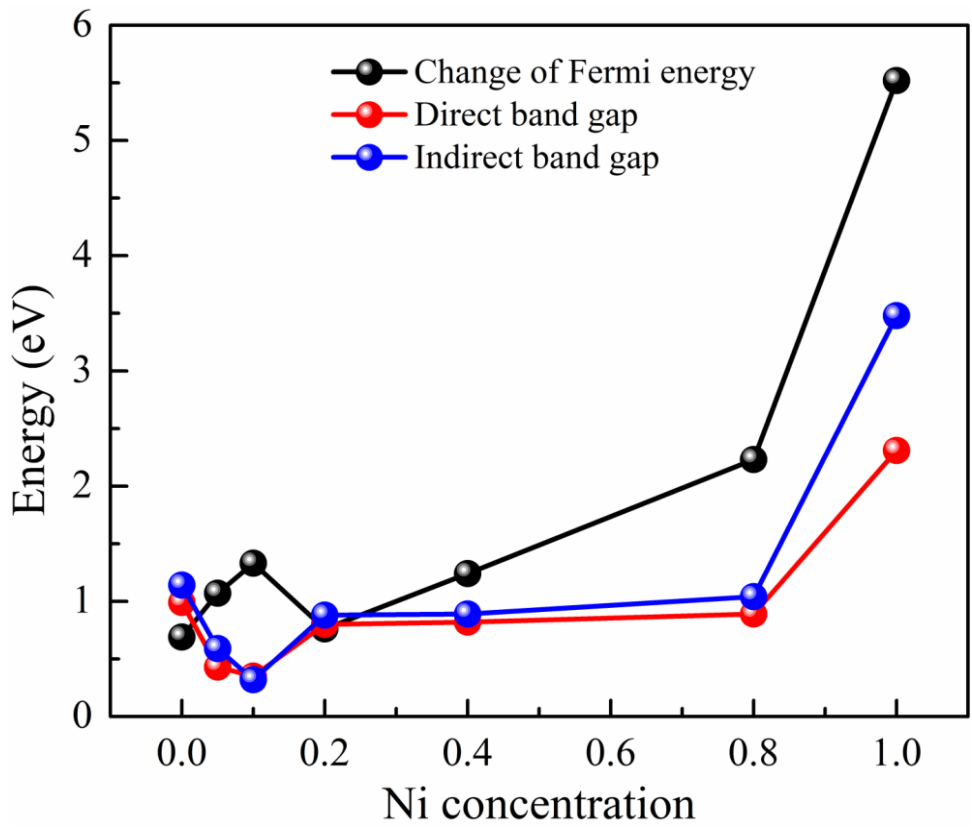

**Figure-15**